\pdfoutput=1
\documentclass[acmsmall]{acmart}
\AtBeginDocument{%
  \providecommand\BibTeX{{%
    \normalfont B\kern-0.5em{\scshape i\kern-0.25em b}\kern-0.8em\TeX}}}

\setcopyright{rightsretained}
\acmJournal{PACMMOD}
\acmYear{2024} \acmVolume{2} \acmNumber{3 (SIGMOD)} \acmArticle{129} \acmMonth{6}\acmDOI{10.1145/3654932}

\makeatletter
\gdef\@copyrightpermission{
 \begin{minipage}{0.2\columnwidth}
  \href{https://creativecommons.org/licenses/by-nc/4.0/}{\includegraphics[width=0.90\textwidth]{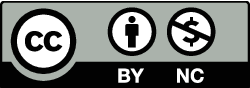} }
 \end{minipage}\hfill
 \begin{minipage}{0.8\columnwidth}
  \href{https://creativecommons.org/licenses/by-nc/4.0/}{This  work is licensed under a Creative Commons Attribution-NonCommercial International 4.0 License.}
 \end{minipage}
 \vspace{5pt}
}
\makeatother

\usepackage{algorithm}
\usepackage{algpseudocode}
\usepackage{listings}
\usepackage{pgf}
\usepackage{import}
\usepackage{tikz}
\usetikzlibrary{calc}
\usetikzlibrary{positioning}
\usetikzlibrary{shapes}
\usetikzlibrary{fit}
\usetikzlibrary{arrows.meta}

\usepackage{enumitem}
\setlist[itemize]{leftmargin=2em}

\usepackage{amsmath}
\usepackage{amsfonts}
\usepackage{amsthm}
\usepackage{mathtools}
\usepackage{bm}

\DeclareMathOperator{\E}{\mathbb{E}}          %
\DeclareMathOperator{\Var}{Var}               %
\DeclareMathOperator{\1}{\mathbf{1}}          %

\DeclarePairedDelimiter\rdbr{\lparen}{\rparen}
\DeclarePairedDelimiter\sqbr{\lbrack}{\rbrack}
\DeclarePairedDelimiter\anbr{\langle}{\rangle}
\DeclarePairedDelimiter\set\{\}
\DeclarePairedDelimiter\abs{\lvert}{\rvert}

\DeclarePairedDelimiter\norm{\lVert}{\rVert}

\def\reals{\mathbb{R}}  %
\def\comps{\mathbb{C}}  %

\def\rvc{{\mathbf{c}}}

\def\ervc{{\textnormal{c}}}

\def\vzero{{\bm{0}}}
\def\vone{{\bm{1}}}

\def\va{{\bm{a}}}

\def\vc{{\bm{c}}}

\def\vf{{\bm{f}}}
\def\vg{{\bm{g}}}

\def\vx{{\bm{x}}}
\def\vy{{\bm{y}}}

\def\evc{{c}}

\def\evf{{f}}

\def\evx{{x}}
\def\evy{{y}}

\def\mPi{{\bm{\Pi}}}

\def\emPi{{\Pi}}

\newcommand{\tens}[1]{\bm{\mathcal{#1}}}

\def\tF{{\tens{F}}}

\newcommand{\etens}[1]{\mathcal{#1}}

\def\etF{{\etens{F}}}

\algnewcommand{\LineComment}[1]{\State \(\triangleright\) #1}

\definecolor{red}{HTML}{E31C23}
\definecolor{green}{HTML}{1AA24A}
\definecolor{blue}{HTML}{005CAD}
\definecolor{purple}{HTML}{724096}
\definecolor{orange}{HTML}{F58B0A}

\newtheorem{definition}{Definition}[section]
\newtheorem{theorem}{Theorem}[section]

\newcommand{\revised}[1]{{#1}}

\begin{document}

\title{Convolution and Cross-Correlation of Count Sketches Enables Fast Cardinality Estimation of Multi-Join Queries}

\author{Mike Heddes}
\orcid{0000-0002-9276-458X}
\affiliation{%
  \institution{University of California, Irvine}
  \city{Irvine}
  \state{CA}
  \country{USA}
}
\email{mheddes@uci.edu}

\author{Igor Nunes}
\orcid{0000-0002-8443-4708}
\affiliation{%
  \institution{University of California, Irvine}
  \city{Irvine}
  \state{CA}
  \country{USA}
}
\email{igord@uci.edu}

\author{Tony Givargis}
\orcid{0000-0002-1608-9324}
\affiliation{%
  \institution{University of California, Irvine}
  \city{Irvine}
  \state{CA}
  \country{USA}
}
\email{givargis@uci.edu}

\author{Alex Nicolau}
\orcid{0009-0003-9833-8455}
\affiliation{%
  \institution{University of California, Irvine}
  \city{Irvine}
  \state{CA}
  \country{USA}
}
\email{anicolau@uci.edu}

\renewcommand{\shortauthors}{Mike Heddes, Igor Nunes, Tony Givargis, \& Alex Nicolau}

\begin{abstract}
With the increasing rate of data generated by critical systems, estimating functions on streaming data has become essential. 
This demand has driven numerous advancements in algorithms designed to efficiently query and analyze one or more data streams while operating under memory constraints. 
The primary challenge arises from the rapid influx of new items, requiring algorithms that enable efficient incremental processing of streams in order to keep up. 
A prominent algorithm in this domain is the \textit{AMS sketch}. Originally developed to estimate the second frequency moment of a data stream, it can also estimate the cardinality of the equi-join between two relations. 
Since then, two important advancements are the \textit{Count sketch}, a method which significantly improves upon the sketch update time, and secondly, an extension of the AMS sketch to accommodate multi-join queries. 
However, combining the strengths of these methods to maintain sketches for multi-join queries while ensuring fast update times is a non-trivial task, and has remained an open problem for decades as highlighted in the existing literature. 
In this work, we successfully address this problem by introducing a novel sketching method which has fast updates, even for sketches capable of accurately estimating the cardinality of complex multi-join queries.
We prove that our estimator is unbiased and has the same error guarantees as the AMS-based method. 
Our experimental results confirm the significant improvement in update time complexity, resulting in orders of magnitude faster estimates, with equal or better estimation accuracy.
\end{abstract}

\ccsdesc[500]{Information systems~Query optimization}
\ccsdesc[500]{Theory of computation~Streaming models}
\ccsdesc[500]{Theory of computation~Sketching and sampling}

\keywords{Cardinality Estimation, Sketching, Synopsis Data Structures}

\received{October 2023}
\received[revised]{January 2024}
\received[accepted]{February 2024}

\maketitle

\section{Introduction}
\label{sec:introduction}

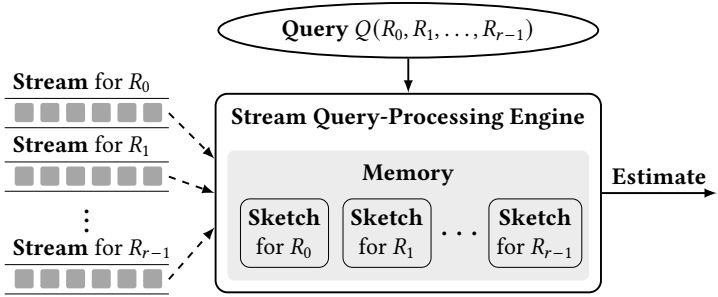
\begin{figure}[ht]
    \centering
    \scalebox{0.9}{\begin{tikzpicture}
    \node[draw=none, thick, minimum width=2.4cm, minimum height=.4cm, align=left] (r1) at (0,0) {};
    \foreach \i in {1,2,...,6} {
        \draw[draw=none ,fill=gray!70, rounded corners=1pt] ([xshift=-\i*0.38cm+.28cm, yshift=-1.9pt]r1.north east) rectangle ++(-0.28cm,-0.28cm);
    }
    \node[above right, yshift=-1pt] at (r1.north west) {\textbf{Stream} for $R_0$};
    \draw[black] (r1.north west) -- (r1.north east);
    \draw[black] (r1.south west) -- (r1.south east);

    \node[draw=none, thick, minimum width=2.4cm, minimum height=.4cm, below=0.5cm of r1, align=left] (r2) {};
    \foreach \i in {1,2,...,6} {
        \draw[draw=none ,fill=gray!70, rounded corners=1pt] ([xshift=-\i*0.38cm+.28cm, yshift=-1.9pt]r2.north east) rectangle ++(-0.28cm,-0.28cm);
    }
    \node[above right, yshift=-1pt] at (r2.north west) {\textbf{Stream} for $R_1$};
    \draw[black] (r2.north west) -- (r2.north east);
    \draw[black] (r2.south west) -- (r2.south east);

    \node [draw=none, below=-2pt of r2] (vdots)
    {\fontsize{15}{15}\selectfont $\vdots$};

    \node[draw=none, thick, minimum width=2.4cm, minimum height=.4cm, below=2cm of r1, align=left] (rl) {};
    \foreach \i in {1,2,...,6} {
        \draw[draw=none ,fill=gray!70, rounded corners=1pt] ([xshift=-\i*0.38cm+.28cm, yshift=-1.9pt]rl.north east) rectangle ++(-0.28cm,-0.28cm);
    }
    \node[above right, yshift=-1pt] at (rl.north west) {\textbf{Stream} for $R_{r-1}$};
    \draw[black] (rl.north west) -- (rl.north east);
    \draw[black] (rl.south west) -- (rl.south east);

    \node[draw=black, fill=none, thick, minimum width=5.7cm, minimum height=2.9cm, rounded corners=5pt, right=1.7cm of vdots, yshift=.3cm] (engine) {};
    \node[below=0.1cm of engine.north,align=center]{\textbf{Stream Query-Processing Engine}};

    \node[draw=none, thick, fill=gray!15, rounded corners=3pt, minimum width=5.35cm, minimum height=1.9cm, below=-2.1cm of engine] (mem) {};
    \node[below=0.1cm of mem.north]{\textbf{Memory}};
    \node[draw=black, fill=none, rounded corners, minimum width=1.3cm, minimum height=1cm, anchor=south west, xshift=2mm, yshift=2mm, align=center] (sk1) at (mem.south west) {\textbf{Sketch}\\for $R_0$};
    \node[draw=black, fill=none, rounded corners, minimum width=1.3cm, minimum height=1cm, right=.2cm of sk1, align=center] (sk2) {\textbf{Sketch}\\for $R_1$};
    \node [draw=none, right=0cm of sk2, align=center] (dots)
    {\fontsize{15}{15}\selectfont $\dots$};
    \node[draw=black, fill=none, rounded corners, minimum width=1.3cm, minimum height=1cm, right=-2pt of dots, align=center] (skl) {\textbf{Sketch}\\for $R_{r-1}$};
    
    \draw[-latex, thick, dashed] (r1.east) -- ([yshift=5mm]engine.west);
    \draw[-latex, thick, dashed] (r2.east) -- ([yshift=0mm]engine.west);
    \draw[-latex, thick, dashed] (rl.east) -- ([yshift=-5mm]engine.west);

    \node[draw, thick, ellipse, above=.5cm of engine] (query) {\textbf{Query $Q(R_0 ,R_1 , \dots, R_{r-1})$}};
    \draw[-latex, thick] (query.south) -- (engine.north);

    \draw[-latex, thick] (engine.east) -- ++(1.7,0) node[midway, above, align=center] {\textbf{Estimate}};   
\end{tikzpicture}}
    \caption{Streaming query-processing scheme}
    \label{fig:streaming_setup}
\end{figure}

The analysis of streaming data has amassed considerable attention, driven by the increasing demand for real-time data processing, and the remarkable advancements in algorithms that enable efficiently querying and analyzing data streams under memory constraints. Streaming data refers to data that is received sequentially and is often too large to be stored in its entirety, hence requiring algorithms that can process the data on-the-fly~\cite{babu2001continuous,gilbert2001surfing}. Efficiently providing answers to queries over streaming data is vital in numerous application environments, including recommendation systems~\cite{huang2015tencentrec,chen2013terec,chen2020star}, smart cities~\cite{giatrakos2017complex,biem2010ibm}, network traffic monitoring~\cite{cormode2003finding,dobra2002processing}, natural language processing~\cite{goyal2012sketch}, and analysis of market data in financial systems~\cite{cao2014interactive,ross2011nonparametric,stonebraker20058}. 
\revised{Our focus on the streaming data setting stems from its generality. Streaming algorithms are not only effective in streaming settings but also seamlessly extend their applicability to non-streaming scenarios.}
In this work, we present a novel approach to the problem of estimating a crucial collection of complex queries within the general streaming data framework depicted in Figure~\ref{fig:streaming_setup} and elaborated upon below.

The rise of data-intensive applications has created a need for data structures that can handle massive volumes of data efficiently. This context motivated the emergence of \textit{synopsis data structures}~\cite{gibbons1999synopsis}, a family of data structures designed to represent large quantities of data with sublinear space complexity, which is imperative in the given context.
Examples include random samples, histograms, wavelets and sketches~\cite{cormode2011synopses}, all of which are actively being researched as a means of analyzing and querying streaming data~\cite{cormode2022current,li2018approximate,cormode2011synopses,aggarwal2007survey}. 
These algorithms operate by generating a compressed representation of the original data, which can then be utilized to estimate a specific property or a set thereof. 
For example, the popular Bloom Filter~\cite{bloom1970space} is widely used for membership testing, while the Count-Min sketch~\cite{cormode2005improved} is commonly used for frequency estimation. Both of these methods are examples of sketches. Besides their supported queries, various factors differentiate sketching methods, including sketch size, sketch initialization time, update time, and inference time~\cite{cormode2011sketch} (see Section~\ref{sec:background} for details). These characteristics serve as catalysts for diverse research avenues and are crucial to consider when utilizing or developing a sketching method that is tailored to a specific use case. %

Aside from basic statistical properties such as count, sum, and mean, much useful information from a data stream is derived from its frequency distribution, or histogram. This becomes particularly relevant when we need to compare or estimate functions across multiple data sets, such as the number of shared items. Frequency-based sketches are a class of sketching methods specifically designed for estimating functions of the frequency vector. Among these, the \textit{AMS} (Alon-Matias-Szegedy) sketch~\cite{alon1996space}, also known as \textit{Tug-of-War} sketch, stands out as a prime example, renowned for its established reputation of being both simple and remarkably effective in a wide array of applications. %
\revised{The AMS sketch was initially introduced to estimate the second frequency moment of a data stream, but it was later demonstrated to also estimate the cardinality of any equi-join between two relations~\citep{alon1999tracking}.}

Interestingly, it turns out that many important functions on the frequency vector can be expressed as the cardinality of an equi-join. This equivalence is an important driver behind the development of sketches, often seen as an approximate query processing (AQP) technique~\citep{li2018approximate}. One particularly relevant use case is estimating the join cardinality, which is crucial for query optimizers to efficiently assess the cost of candidate physical join plans. The challenge of determining an appropriate join order is a highly researched problem in the field of databases~\cite{chaudhuri1998overview,lan2021survey}, and the methods employed typically rely on cardinality estimates as the primary input~\cite{leis2015good}. %

Two significant breakthroughs emerged a few years after the introduction of the AMS sketch. First, \citet{charikar2002finding} proposed the \textit{Count sketch}, which divides estimates into ``buckets'' instead of computing the mean of multiple independent and identically distributed (i.i.d.) estimates. This approach makes the sketch more accurate for skewed data and dramatically speeds up its updates~\cite{rusu2008sketches, thorup2004tabulation}. 
Second, \citet{dobra2002processing} proposed a generalization of the AMS sketch that enables the cardinality estimation of multi-join queries, thus considerably expanding the algorithm's applicability.

Although both methods have gained popularity for their respective advantages, the existing literature has highlighted the task of integrating all these benefits into a unified approach as a challenging and unresolved problem~\cite{cormode2011sketch, izenov2021compass}. The specific challenge lies in effectively handling multi-join queries with fast updates. This challenge becomes even more significant when considering the prevalence of such multi-joins, as they constitute the majority of queries (see Section~\ref{sec:databases}). %
The difficulty of combining the Count sketch with the AMS-based multi-join query estimation method arises from the use of binning, as we will discuss in Section~\ref{sec:method}, yet binning is essential for achieving the benefits of the Count sketch. 

\revised{To address this, we propose a new %
sketch that combines insights from both \citet{charikar2002finding} and \citet{dobra2002processing}.
The proposed method relies on the intuitive observation that the operation used to merge single-item AMS sketches to form sketches of tuples, the Hadamard product, is incongruous with the sparse nature of the Count sketch. 
In essence, when two Count sketches undergo the Hadamard product, the resulting sketch will likely lose information due to the sparsity of the Count sketches.

The core innovation of our approach lies in employing circular convolution instead of the Hadamard product for counting tuples in a data stream. We show that, unlike the Hadamard product, this operation ensures the preservation of information from the operands in the resulting Count sketch. This is complemented by incorporating circular cross-correlation in the estimation procedure. Our method not only exhibits superior estimation accuracy when applied to real data and queries, but also operates within the same memory constraints. Moreover, we have significantly improved the time complexity of the sketch update process, enabling estimates to be computed orders of magnitude faster.}
We prove that our estimator is unbiased and offers error guarantees equivalent to \citet{dobra2002processing}.
Importantly, our method does not require prior knowledge of the data distribution. Our empirical findings support the practical applicability of the proposed method, underscoring its significant advancement in addressing the aforementioned open problem.

\section{Background}
\label{sec:background}

This section provides the necessary background and introduces the key methodologies and notation used in this work. For an overview of the notation see Table~\ref{tab:notation}. 
These concepts and methodologies set the foundation for the introduction of our proposed method.

\begin{table}[h]
    \centering
    \caption{\label{tab:notation} Notation}
    \begin{tabular}{l|l}
        \toprule
        Symbol & Definition \\
        \midrule
        $\sqbr{n} = \set{0, 1, \dots, n - 1}$ & Domain of items\\
        $(i, \Delta)$ & Tuple of item and frequency change\\
        $\vf, \vg \in \reals^{n}$ & Frequency vectors\\
        $\mPi \in \reals^{m\times n}$ & Random matrix\\
        $\vc \in \reals^m$ & Vector of counters, i.e., the sketch\\
        $s_j: \sqbr{n} \to \set{-1, +1}$ & Random sign function\\
        $h_j: \sqbr{n} \to \sqbr{m}$ & Random bin function\\
        $R_0, R_1, \dots, R_{r-1}$ & Database relations\\
        $Q(R_0, R_1, \dots, R_{r-1})$ & Query over relations\\
        $u, v\in \sqbr{w}$ & Joined attribute names (vertices)\\
        $\set{u, v} \in E$ & Join from all joins (edges)\\
        $u \in \Omega(R_k)$ & Joined attribute $u$ of relation $R_k$\\
        $v \in \Gamma(u)$ & Joined attribute $v$ with $u$\\
        $\Psi(u) \in \sqbr{w-r + 1}$ & Join graph component of attribute $u$\\
        $I_k = \sqbr{n}\times\cdots\times\sqbr{n}$ & Domain of relation $R_k$\\
        $\etF_k(i)$ & Frequency of tuple $i$ in relation $R_k$\\
        $X, \E\sqbr{X}$ & Estimate and expected value\\
        $\epsilon, \delta$ & Error bound and confidence\\
        $y, \hat{y}$ & True and predicted cardinality\\
        \bottomrule
    \end{tabular}
\end{table}

\subsection{Streaming data}
\label{sec:data-stream}

The focus of this paper is on \textit{streaming data} analysis, a prominent application area for synopsis data structures~\cite{gibbons1999synopsis}, which involves real-time processing of data that arrives at a high frequency. 
Streaming data naturally arises in many big data applications, including network traffic monitoring~\cite{cormode2003finding,dobra2002processing}, recommendation systems~\cite{huang2015tencentrec,chen2013terec,chen2020star}, natural language processing~\cite{goyal2012sketch}, smart cities~\cite{giatrakos2017complex,biem2010ibm}, and analysis of market data in financial systems~\cite{cao2014interactive,ross2011nonparametric,stonebraker20058}. 
Algorithms for streaming data are designed to handle data that can only be observed once, in arbitrary order, as it continuously arrives~\cite{dobra2002processing}. 
Consequently, these algorithms must be highly efficient in processing each input, while utilizing limited memory resources, to keep up with the rapid influx of new data.

\revised{
By presenting our method within the streaming data setting, we establish its applicability to a broad range of scenarios.
This is because streaming algorithms are also applicable when multiple data accesses or a specific access order are allowed. Inversely, algorithms that require multiple data accesses or a specific access order, like many learning-based methods, clearly do not apply in a streaming data setting.
Even when a streaming algorithm is not strictly necessary, optimizing for fewer data accesses remains advantageous because it minimizes potentially costly I/O operations. 
}

We formulate the problem as follows, based on \citet{cormode2005improved}: consider a vector $\vf(t) \in \reals^n$, which is assumed too large to be stored explicitly and is therefore presented implicitly in an incremental fashion. 
Starting as a zero vector, $\vf(t)$ is updated by a stream of pairs $(i_t, \Delta_t)$ which increments the $i_t$-th element by $\Delta_t$, meaning that $\evf_{i_t}( t) = \evf_{i_t}(t-1) + \Delta_t$, while the other dimensions remain unchanged. The \textit{items} $i_t$ are members of the \textit{domain}\footnote{Without loss of generality, we can assume $i\in\sqbr{n}$~\citep{dobra2002processing, ganguly2004tracking}.} $\sqbr{n} = \set{0,1,\dots,n-1}$; $\Delta_t\in\reals$ are the \textit{changes in frequency} %
and $\vf(t)$ is called the \textit{frequency vector}. At any time $t$, a \textit{query} may request the computation of a function on $\vf(t)$. 
Specific streaming settings are further classified by their type of updates, as follows:
\begin{itemize}
    \item \textbf{cash-register:} $\Delta_t > 0$ on every update;
    \item \textbf{strict turnstile:} for some updates $\Delta_t$ can be negative, but $\evf_i(t) \geq 0$ for all $i$ and $t$;
    \item \textbf{general turnstile:} both updates and entries of the vector $\vf(t)$ can assume negative values at any time $t>0$.
\end{itemize}

The algorithms we discuss utilize synopsis data structures to efficiently handle data streams, eliminating the need to explicitly store and compute over $\vf(t)$ to answer a set of supported queries. 
In the following section, we will introduce a group of sketching techniques known as \textit{linear sketches}. This family of methods, which includes the approach proposed in this paper, is designed to support the most general streaming setting, i.e., the general turnstile. 

\subsection{Linear sketching}
\label{sec:frequency-sketches}

Sketching techniques are a popular set of methods for dealing with streaming data and approximate query processing~\cite{cormode2011sketch}. 
Both the baselines and the method proposed in this paper are \textit{linear} sketches, meaning that the summaries they generate can be represented as a \textit{linear transformation} of the input. In contrast, the Bloom filter~\cite{bloom1970space} serves as a classic example of a \textit{non}-linear sketch. 

Formally, for a given vector $\vx \in \reals^n$, we define a linear sketch as a vector obtained by $\mPi \vx$, where $\mPi \in \reals^{m\times n}$ is some random matrix, and $m\ll n$. 
The linearity of the transformation offers notable advantages~\citep{cormode2011sketch}: it allows for processing items in any order and combining different sketches through addition. This enables efficient handling of data and supports map-reduce style processing of large data streams. 
In every sketching method, the random matrix is thoughtfully designed to enable the estimation of one or multiple functions over $\vx$, utilizing only its ``summary'' captured by $\mPi\vx$, thereby eliminating the necessity for accessing $\vx$ itself. 
When sketching techniques are used in a streaming scenario they are often referred to as \textit{frequency-based sketches}, where the input vector $\vx$ is the frequency vector $\vf(t)$ defined in Section~\ref{sec:data-stream}. Hereafter, we will omit the time argument from the frequency vector for brevity.

At this point, one naturally wonders: how can we transform the vector $\vf$ of size $n$, which is already considered too large, using a matrix $\mPi$ that is even larger with size $mn$? Streaming algorithms cleverly represent the matrix $\mPi$ succinctly using hash functions, enabling them to generate just the column of $\mPi$ that is needed to add a given item. Further details on this process will be provided later. Furthermore, it is crucial to ensure that the sketch is efficiently updated as $\vf$ changes, i.e, as new items stream in. This can be achieved by making $\mPi$ sparse, as we will explore shortly. The effectiveness and versatility of a sketching method primarily relies on the following key properties~\cite{cormode2011sketch}:

\begin{itemize}
    \item \textbf{Sketch size:} The total number of counters and random seeds required by the sketch, determined by the parameter $m$. 
    \item \textbf{Initialization time:} The time it takes to initialize the sketches. Typically, this involves simply setting a block of memory to zeros, and sampling the random seeds for the hash functions.
    \item \textbf{Update time:} In streaming settings, algorithms must keep pace with the high influx of items. The update time determines the highest item throughput rate that can be sustained.  
    \item \textbf{Inference time:} The time it takes to compute an estimate from the generated sketches.
    \item \textbf{Accuracy:} It is crucial to understand the accuracy of an estimator for a given memory budget (limiting the sketch size) and throughput requirement (constraining the update time).
    \item \textbf{Supported queries:} Each sketch is designed to enable estimation of a specific set of functions on the input vector. Typically, a query-specific procedure needs to be performed on the sketch to approximate the value of a particular function.
\end{itemize}
In the following, we will describe some important sketching methods from the existing literature that have particularly space- and time-efficient ways of representing and computing $\mPi\vf$.

\subsection{AMS sketch}
\label{sec:ams}
The \textit{AMS} sketch, also referred to as the \textit{Tug-of-War} or \textit{AGMS} sketch, is a pioneering technique for frequency-based sketching that was first introduced by Alon, Matias, and Szegedy (AMS)~\cite{alon1996space}. The method was originally proposed as a way to estimate the \textit{second frequency moment} $F_2$ of a data stream, where $F_2 = \norm{\vf}^2_2 = \sum_{i=0}^{n-1} \evf_{i}^2$ and $\vf$ is the frequency vector of the stream as defined in Section~\ref{sec:data-stream}. 
The AMS sketch is represented by a vector $\rvc$, containing $m = O(1/\epsilon^{2})$ counters $\ervc_j$, for $j\in\sqbr{m}$, where $0 < \epsilon < 1$ is the relative error bound. %
The counters are i.i.d. random variables obtained by $\ervc_j=\sum_{i=0}^{n-1} \evf_i s_j(i)$, where each $s_j: \sqbr{n} \to \set{-1, +1}$ is drawn from a family of 4-wise independent hash functions (see Definition~\ref{def:kwise-hash}). These hash functions are used to compute the random projection $\mPi\vf$ without representing $\mPi$ explicitly, since $\emPi_{j,i} = s_j(i)$.
To establish a confidence level $\delta$, one can take the median of $O\rdbr{\log 1/\delta}$ independent estimates. The overall method then requires only $O\rdbr{\rdbr{1/\epsilon^{2}} \log 1/\delta}$ counters. 
Taking the median of i.i.d. estimates, sometimes called the ``median trick'', is universal among sketching methods because it is an effective way to rapidly improve the confidence level of an estimate by the Chernoff Bound~\citep{alon1996space}. We will, therefore, revisit this concept in the subsequent discussions of other methods.

\begin{definition}[$k$-wise independence~\citep{wegman1981new, pagh2013compressed}]
\label{def:kwise-hash}
A family of hash functions $H = \set{h: \sqbr{n} \to \sqbr{m}}$ is said to be $k$-wise independent if for any $k$ distinct items $x_0, \dots, x_{k-1}$ the hashed values $h(x_0), \dots, h(x_{k-1})$ are independent and uniformly distributed in $\sqbr{m}$.
\end{definition}

In their original work, \citet{alon1996space} showed that $\frac{1}{m} \anbr{\mPi\vf, \mPi\vf}$ is an unbiased estimator of $F_2$. 
In fact, it can be demonstrated more generally that for any two vectors $\vf$ and $\vg$, the normalized inner product of their AMS sketches $\frac{1}{m}\anbr{\mPi\vf, \mPi\vg}$ is an unbiased estimator for their inner product $\anbr{\vf, \vg}$~\cite{alon1999tracking}. 
Notably, when $\vf$ and $\vg$ correspond to the frequency vectors of a given attribute of two database relations, this estimated value corresponds to the equi-join size of these relations over that attribute. Theorem~\ref{thm:ams} formally states the expectation, and bounds the variance of the AMS sketch.

\begin{theorem}[AMS sketch]
\label{thm:ams}
For any vectors $\vf, \vg \in \reals^n$ and a random matrix $\mPi \in \reals^{m\times n}$ constructed by 4-wise independent hash functions $s_j\colon\sqbr{n}\to\set{-1, +1}$ for $j \in \sqbr{m}$ and $\emPi_{j,i} = s_j(i)$, we have:
\begin{align*}
    \E\sqbr*{\frac{\anbr{\mPi\vf, \mPi\vg}}{m}} = \anbr{\vf, \vg},\quad\mathrm{and}\quad\Var\rdbr*{\frac{\anbr{\mPi\vf, \mPi\vg}}{m}} \leq \frac{2}{m} \norm{\vf}^2_2\norm{\vg}^2_2
\end{align*}
\end{theorem}

\begin{proof}
    See Lemma 4.4 of \citet{alon1999tracking}.
\end{proof}

\begin{figure}
    \centering
    \scalebox{0.9}{\begin{tikzpicture}[every node/.style={inner sep=0pt}]

\node[align=left] at (3.5,0.0) (I) {$(i, \Delta)$};

\foreach \i in {0,1,2} {
    \node[draw=none, fill=gray!70, minimum size=0.5cm, rounded corners=2pt] at (\i*0.6, 0.0) (I\i) {};
}
\draw[solid] (-0.4, -0.35) -- (1.6, -0.35);
\draw[solid] (-0.4, 0.35) -- (1.6, 0.35);

\draw[-latex, thick, solid] ([xshift=0.2cm]I2.east) -- ([xshift=-0.2cm]I.west);
\node[] at (0.1, 0.6) {\textbf{Stream}};

\foreach \i in {0,1,2,3} {
    \pgfmathsetmacro{\x}{1cm + \pgflinewidth}
    
    \node[draw=none, minimum height=0.7cm, minimum width=1.0cm] at (\i*\x+1.2cm, 1.6) (A\i) {$\evc_\i$};
    \draw[-latex, thick, solid] ([xshift=0mm,yshift=1mm]I.north) -- ([yshift=-1mm]A\i.south);
    
    \node[draw=none, minimum height=0.7cm, minimum width=1.0cm] at (\i*\x+1.2cm, -1.6) (C\i) {$\evc_\i$};
}

\node[draw=none, minimum height=0.7cm, minimum width=1.0cm, right=1cm of A3.west] (AE) {$\cdots$};
\node[draw=none, minimum height=0.7cm, minimum width=1.0cm, right=1cm of C3.west] (CE) {$\cdots$};

\node[draw, thick, minimum height=0.7cm, minimum width=1cm, right=1cm of AE.west] (AM) {$\evc_{m-1}$};
\node[draw, thick, minimum height=0.7cm, minimum width=1cm, right=1cm of CE.west] (CM) {$\evc_{m-1}$};

\foreach \i in {0,1,2} {
    \draw[thick, solid] (A\i.north east) -- (A\i.south east);
    \draw[thick, solid] (C\i.north east) -- (C\i.south east);
}

\draw[thick, solid] (A0.north west) -- (A3.north east) -- (A3.south east) -- (A0.south west) -- (A0.north west);

\draw[thick, solid] (C0.north west) -- (C3.north east) -- (C3.south east) -- (C0.south west) -- (C0.north west);

\node[] at (4.7,0.9) {$\dots$};
\draw[-latex, thick, solid] ([xshift=0mm,yshift=1mm]I.north) -- ([yshift=-1mm]AM.south);

\draw[-latex, thick, solid] ([xshift=0mm,yshift=-1mm]I.south) -- ([yshift=1mm]C2.north) node [midway,xshift=-1.5em] {$h(i)$};

\node[above right, align=left, yshift=0.1cm] at (A0.north west) {\textbf{AMS}};
\node[above right, align=left, yshift=0.1cm] at (C0.north west) {\textbf{Count}};

\end{tikzpicture}}
    \caption{Comparison of the AMS and Count sketches performing a sketch update for an item in the stream.}
    \label{fig:ams-vs-count}
\end{figure}
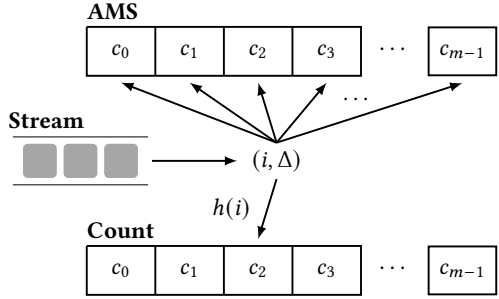

\subsection{Count sketch}
\label{sec:fast-ams-sketch}

In order to ensure fast sketch updates, i.e., sublinear with respect to its size $m$, it is desirable for sketching methods to use a \textit{sparse} linear transformation $\mPi$ when processing input vectors. However, note that the AMS sketch requires changes in all $m$ counters for each update. Consequently, the accuracy of the estimator is constrained by the need to maintain a throughput that corresponds to the rate of incoming items as well as the available memory budget.

The \textit{Count sketch} is another linear sketching method that emerged after AMS and overcomes this limitation by allowing the update time to be independent of the sketch size. It achieves this by employing a technique called the ``hashing trick,''~\cite{weinberger2009feature, cormode2011sketch} which ensures that only one counter per estimate is modified during each update. 
As a result, the Count sketch improves the update time complexity from $O\rdbr{\rdbr{1/\epsilon^{2}} \log 1/\delta}$ to just $O\rdbr{\log 1/\delta}$, while maintaining not only the same error bounds, but also the same space and inference time complexities as the AMS sketch.
The AMS sketch and Count sketch update procedures are compared in Figure~\ref{fig:ams-vs-count}.
Although the Count sketch was originally introduced for the \textit{heavy hitters} problem~\citep{charikar2002finding}, the same hashing trick can be applied to speed up the AMS sketch, which is commonly known as the \textit{Fast-AMS sketch}~\citep{cormode2005sketching}. 

The value of each counter in the Count sketch is given by $\ervc_j = \sum_{i=0:h(i) = j}^{n-1} \evf_i s(i)$, where $h: \sqbr{n} \to \sqbr{m}$ is a random bin function drawn from a family of 2-wise independent hash functions, and $s: \sqbr{n} \to \set{-1, +1}$ is again a random sign function drawn from a family of 4-wise independent hash functions. The corresponding random matrix $\mPi$ is specified by $\emPi_{j,i} = s(i)\1(h(i)=j)$. Notice that $\mPi$ has only one non-zero value per column, making it highly sparse. Also, the Count sketch requires only one sign and bin hash function per estimate, regardless of the required precision, resulting in a further reduction in memory usage. An estimate is obtained by taking the inner product between sketches. Theorem~\ref{thm:count-sketch} formally states the expectation and bounds the variance of the Count sketch.

\begin{theorem}[Count sketch]
\label{thm:count-sketch}
For any vectors $\vf, \vg \in \reals^n$ and a random matrix $\mPi \in \reals^{m\times n}$ constructed by 4-wise independent hash function $s\colon\sqbr{n}\to\set{-1, +1}$ and 2-wise independent hash function $h\colon\sqbr{n}\to\sqbr{m}$ with $\emPi_{j,i} = s(i)\1(h(i) = j)$, we have:
\begin{align*}
    \E\sqbr*{\anbr{\mPi\vf, \mPi\vg}} = \anbr{\vf, \vg}, \quad\mathrm{and}\quad\Var\rdbr*{\anbr{\mPi\vf, \mPi\vg}} \leq \frac{2}{m} \norm{\vf}^2_2\norm{\vg}^2_2
\end{align*}
\end{theorem}

\begin{proof}
    See Appendix 22 of \citet{weinberger2009feature} and Lemma 4.4 of \citet{alon1999tracking}.
\end{proof}

The Count sketch has also been shown to outperform the AMS sketch in estimation precision for skewed data distributions~\citep{rusu2008sketches}. This is because the Count sketch is able to separate out the few high frequency components with high probability. This is an important trait, as it has been widely acknowledged in the literature that the majority of real-world data distributions exhibit a skewed nature~\citep{dobra2002processing, leis2015good, manerikar2009frequent,yang2017pyramid,roy2016augmented}. We further discuss this topic in Section~\ref{sec:databases}.

\subsection{Extensions to the Count sketch}
\label{sec:tensor-sketch}

The Count sketch was originally introduced for single-dimensional data which is represented by the frequency vector $\vf$. More recently, however, several extensions to the Count sketch have been proposed which enable its usage for higher-dimensional data represented by a frequency \textit{tensor} $\tF$ instead~\citep{liu2022tensor}. The most notable extensions are the \textit{Tensor sketch}~\citep{pham2013fast} and the \textit{Higher-Order Count (HOC) sketch}~\citep{shi2019higher}.
Both methods set out to reduce the computational complexity of machine learning applications. The Tensor sketch was used to approximate the polynomial kernel, but finds its origin in estimating matrix multiplication~\citep{pagh2013compressed}. The HOC sketch was introduced to compress the training data or neural network parameters, in order to speed up training and inference processes.

For each incoming item of the stream, both methods start by encoding all axes separately using independent instances of the Count sketch. They differ in the way these individually sketched axes are combined: the Tensor sketch employs circular convolution \revised{(see Definition~\ref{def:circ-conv})}, generating a sketch vector, whereas the HOC sketch utilizes the tensor product to produce a sketch tensor of the same order but with reduced dimensions compared to $\tF$.
The use of the tensor product ensures that axis information about the sketched data is preserved, at the cost of an exponential increase in sketch size with the order of the tensor. The circular convolution, in contrast, preserves the dimensionality of the sketch vector for any tensor order, i.e., it maps tensors to vectors. 

In the context of databases, COMPASS~\cite{izenov2021compass} uses HOC sketches to estimate the cardinality of multi-join queries. They additionally propose a method to approximate HOC sketches by merging Count sketches. While they show promising results for query optimization, their estimation method lacks theoretical error guarantees. Moreover, Section~\ref{sec:exp-accuracy} will show that, in practice, our proposed method achieves significantly higher estimation accuracy.
The Tensor sketch (see Definition~\ref{def:tensor-sketch}) is the most related to our method, however, our method and application are novel and solves an important open problem in the streaming and databases community, as will be detailed in Section~\ref{sec:method}. 

\revised{
\begin{definition}[Circular convolution]
\label{def:circ-conv}
The circular convolution $\vx * \vy$ of any two vectors $\vx, \vy \in \comps^m$ is a vector with elements given by $(\vx * \vy)_j = \sum_{i=0}^{m-1} \evx_i \evy_{(j-i) \bmod m}$ for all $j \in \sqbr{m}$.
\end{definition}
}

\begin{definition}[Tensor sketch~\citep{pagh2013compressed, pham2013fast}]
\label{def:tensor-sketch}
Consider any order $d$ tensor $\tF \in \reals^{n^d}$ and random matrix $\mPi \in \reals^{m\times n^d}$ constructed by 2-wise independent hash functions $h_k \colon \sqbr{n} \to \sqbr{m}$ and 4-wise independent hash functions $s_k \colon \sqbr{n} \to \set{-1, +1}$ for each axis $k \in \sqbr{d}$. Let $\emPi_{j, i} = S(i)\1(H(i)=j)$ with the following decomposable hash functions:
\begin{gather*}
    H(i) = \rdbr*{\sum_{k=0}^{d-1} h_k(i_k)} \bmod{m},\quad S(i) = \prod_{k=0}^{d-1} s_k(i_k)
\end{gather*}
Then, the Tensor sketch is given by $\mPi\tF$.
\end{definition}

\subsection{Multi-join with AMS sketches}
\label{sec:AMS-multijoin}

\begin{figure}
    \centering
    \begin{tikzpicture}
    \tikzset{every node/.append style={draw, line width=1pt}}

    \node[minimum height=0.7cm, circle, draw=blue] (A) at (0,0) {$0$};
    \node[minimum height=0.7cm, circle, draw=blue, below=3em of A] (B) {$3$};
    \node[minimum height=0.7cm, circle, draw=blue, below right=1.3em and 3.5em of A] (C) {$1$};

    \node[minimum height=0.7cm, circle, draw=red, right=.5em of C] (D) {$2$};
    \node[minimum height=0.7cm, circle, draw=red, right=3em of D] (E) {$4$};

    \draw[minimum height=0.7cm, blue, line width=1pt] (B) -- (C);
    \draw[minimum height=0.7cm, blue, line width=1pt] (C) -- (A);
    \draw[minimum height=0.7cm, red, line width=1pt] (D) -- (E);

    \pgfdeclarelayer{bg} %
    \pgfsetlayers{bg,main} %
    \begin{pgfonlayer}{bg}
    \node[fit=(C)(D), rounded corners, draw=black!10, fill=black!10 , inner sep=.8em, minimum height=1.5cm, yshift=0.1cm, label={[above=-1.5em, font=\small]$R_1$}] {};
    \node[fit=(A), rounded corners, draw=black!10, fill=black!10 , inner sep=.8em, minimum height=1.5cm, yshift=0.1cm, label={[yshift=-.8em, left=.3em, font=\small]$R_0$}] {};
    \node[fit=(B), rounded corners, draw=black!10, fill=black!10 , inner sep=.8em, minimum height=1.5cm, yshift=0.1cm, label={[yshift=-.8em, left=.3em, font=\small]$R_2$}] {};
    \node[fit=(E), rounded corners, draw=black!10, fill=black!10 , inner sep=.8em, minimum height=1.5cm, yshift=0.1cm, label={[yshift=-.8em, left=.3em, font=\small]$R_3$}] {};
    \end{pgfonlayer}
    
\end{tikzpicture}
    \vspace{1em}
\begin{verbatim}
SELECT COUNT(*) FROM R0, R1, R2, R3
WHERE R0.0 = R1.1 AND R2.3 = R1.1 AND R3.4 = R1.2
\end{verbatim}  
    \caption{Example join graph and corresponding SQL query. Additional attributes in each relation, not involved in the join, are omitted for clarity.}
    \label{fig:example-graph-query}
\end{figure}

Another significant advancement in linear sketching techniques emerged around the same period as the Count sketch. \citet{dobra2002processing} proposed a generalization of the AMS sketch that enables the estimation of complex multi-join aggregate queries, such as count and sum queries. These estimates are useful for big data analytics and query optimization~\citep{leis2015good}. The proposed method addresses the scenario where a query $Q(R_0, R_1, \dots, R_{r-1})$ involves multiple relations $R_k$ for $k \in \sqbr{r}$. An example is illustrated in Figure~\ref{fig:example-graph-query}, which provides an intuitive visualization of this type of complex query as a disconnected, undirected graph. In this abstraction, each vertex corresponds to an attribute, edges represent the joins between them, and attributes are grouped to form relations.

The technique generates sketches for each relation by iterating over all the \textit{tuples} $i$ in each relation once. This iterative traversal of the tuples aligns with the streaming data scenario described in Section~\ref{sec:data-stream}, enabling the sketches to be created on-the-fly as the relations are updated. The sketch $\vc_k$ for relation $R_k$ is given by:
\begin{align}
    \ervc_{k,j} = \sum_{i\in I_k} \etF_{k}(i) \prod_{u \in \Omega(R_k)} \prod_{v \in \Gamma(u)} s_{j, \set{u, v}}\rdbr{i_u} \label{eq:dobra}
\end{align}
where we use the following notation: $i$ represents a tuple that belongs to the domain $I_k$ of relation $R_k$; $I_k$ is the cross product of item domains $\sqbr{n} \times \cdots \times \sqbr{n}$ for each joined attribute of relation $R_k$; $\etF_k(i)$ gives the frequency of tuple $i$ in relation $R_k$; $i_u$ denotes the value in tuple $i$ for attribute $u$; $u$ is an attribute from the set of joined attributes of $R_k$ in the query, denoted as $\Omega(R_k)$ (for example, $\Omega(R_1) = \set{1, 2}$ in Figure~\ref{fig:example-graph-query}); $v$ is an attribute from the set of attributes joined with $u$, denoted as $\Gamma(u)$ (for example, $\Gamma(1) = \set{0, 3}$ in Figure~\ref{fig:example-graph-query}). We represent a join between two attributes with $\set{u, v}$. Both $u, v \in \sqbr{w}$ are from the set of all joined attributes. Our notation assumes that all attributes are globally unique, which can easily be achieved in practice, for instance, by concatenating the relation and attribute names. Moreover, following \citet{dobra2002processing}, we assume that joins are non-cyclic, a self-join is thus represented as a join with a fictitious copy of the relation. It is worth noting that the copy does not need to be physically created, this is done solely to simplify the notation. 
Note also that the functions $\Gamma$ and $\Omega$ are defined for a specific query $Q$. We omit this dependence in the notation for brevity, as it is evident from their definitions.

Once the sketches are created, a query estimate is derived by performing the element-wise multiplication of the sketches, often referred to as the \textit{Hadamard product} of sketches.
This is followed by calculating the mean over the counters. Formally, this can be expressed as: $X = \frac{1}{m}\sum_{j=0}^{m-1} \prod_{k=0}^{r-1} \evc_{k,j}$, where $X$ is an unbiased estimate of the cardinality of query $Q$. The expectation and variance of $X$ are formally stated in Theorem~\ref{thm:dobra}.

\begin{theorem}
\label{thm:dobra}
Given an acyclic query of relations $R_k$ for $k \in \sqbr{r}$, let Equation~\ref{eq:dobra} provide the sketches for each relation and $X = \frac{1}{m}\sum_{j=0}^{m-1} \prod_{k=0}^{r-1} \evc_{k,j}$ the cardinality estimate of the query, then we have: 
\begin{align*}
    \E\sqbr{X} &= \sum_{i\in I_0 \times \cdots \times I_{r-1}}  \etF_{0}(i) \cdots \etF_{r-1}(i) \prod_{\set{u,v} \in E} \1(i_u = i_v)\\
    \Var(X) &\leq \frac{1}{m}3^{r-1} \prod_{k=0}^{r-1} \norm*{\tF_k}^2_2
\end{align*}
\end{theorem}

\begin{proof}
In Lemmas 3.1 and 3.2 of \citet{dobra2002processing} similar results are presented, albeit with a slightly looser bound on the variance. However, we were unable to locate the proof for their claims. Therefore, we provide the proof for the presented theorem in Appendix~\ref{sec:proofs} for the sake of completeness.
\end{proof}

From the perspective of the Count sketch extensions discussed in Section~\ref{sec:tensor-sketch}, this method can be interpreted as a similar generalization but for the AMS sketch. It can thus be seen as one of the first methods to generalize sketching for tensor data, although this aspect was not explicitly mentioned in the original work.

\subsection{Other related work}
\label{sec:other}

In this section, we discuss other recent work in cardinality estimation.
The Pessimistic Estimator~\citep{cai2019pessimistic} is an interesting sketching technique which provides an upper bound of the cardinality. They show that it improves upon the cardinality estimator within PostgreSQL. However, the practical use of the method is limited due to its lengthy estimation time~\cite{han2021cardinality}, at times exceeding the query execution time, as also mentioned by the authors.

Since the inception of sketching, a number of techniques have been proposed that complement the aforementioned sketches. Among these techniques, the Augmented Sketch~\citep{roy2016augmented} and the JoinSketch~\citep{wang2023joinsketch} aim to improve the accuracy of sketches for skewed data by separating the high- from the low-frequency items in the sketch. The counters of the high-frequency items are explicitly represented in an additional data structure, thereby preventing them from causing high estimation error due to hash collisions.
Another notable technique is the Pyramid sketch~\citep{yang2017pyramid}, which employs a specialized data structure that dynamically adjusts the number of allocated bits for each counter, preventing overflows in the case of high-frequency items. 
It is important to note that these techniques are proposed as complementary tools, compatible with a variety of sketching methods, including the one proposed in this work.

Recently, there has been a parallel effort aimed at harnessing the power of machine learning for cardinality estimation. Among the various approaches, the most promising ones are data-driven methods that build query-independent models to estimate the joint probability of tuples~\citep{han2021cardinality}. Notable examples of such techniques include DeepDB~\citep{hilprecht13deepdb}, BayesCard~\citep{wu2020bayescard}, NeuroCard~\citep{yang2020neurocard}, and FLAT~\citep{zhu2021flat}. While machine learning-based cardinality estimation techniques have been receiving increasing attention, they still face important limitations: these methods are presently viable only in scenarios where supervised training is feasible, their accuracy has not consistently lived up to expectations~\citep{muller2022selected}, and they prove impractical in situations with frequent data updates, such as streaming settings, due to their high cost of model updates~\citep{han2021cardinality}. In Section~\ref{sec:q-error}, we demonstrate that our proposed method not only avoids these performance limitations but also achieves significantly higher accuracy compared to the machine learning techniques.

\section{Method}
\label{sec:method}

In this section, we present our method to solve the longstanding challenge of integrating the key advantages of the Count sketch, such as its efficient update mechanism and superior accuracy when handling skewed data, into a method that effectively estimates the cardinality of multi-join queries.
Devising such a method is recognized as a challenging task, as underscored not only by the author of the Count-Min sketch~\cite{cormode2011sketch} but also by other recent work in the field~\cite{izenov2021compass}. This acknowledgment highlights the importance of our contribution. The importance of solving this problem is further underscored when considering the prevalence of multi-join queries. In fact, an analysis of two sets of widely recognized benchmarking queries, as discussed in Section~\ref{sec:databases}, reveals that multi-joins constitute approximately 97\% of all their queries~\cite{leis2015good,han2021cardinality}.

\subsection{Key insight for preserving information}

We start with an intuitive discussion on the main insight behind the proposed method, using the example illustrated in Figure~\ref{fig:hada-vs-conv}. The main challenge of combining the Count sketch with the method proposed by \citet{dobra2002processing} lies in the inability to effectively merge Count sketches using the Hadamard product.
\citet{dobra2002processing} create the sketch for a tuple as the Hadamard product of the AMS sketches for each value in the tuple. 
As discussed earlier, the advantages of the Count sketch stem from its sparsity. However, as illustrated in the left part of the figure, it is precisely this characteristic that results in a loss of information, with high probability, when combining sketches with the Hadamard product. Essentially, due to the sparsity, the non-zero entry in each sketch is highly likely to appear at a different position, causing the result of the element-wise multiplication to yield a zero vector, devoid of information.

To address this issue, the core concept behind our method, as depicted in the right part of Figure 4, involves the utilization of circular convolution paired with circular cross-correlation \revised{(see Definitions \ref{def:circ-conv} and \ref{def:circ-cross})} during inference. With circular convolution, the resulting sketch has the product of the non-zero entries in the bin that corresponds to the sum of the non-zero indices, modulo $m$. Unlike the Hadamard product, this operation guarantees that the information is preserved. We will delve deeper into this intuition and formalize it in the subsequent sections. \revised{Crucially, the circular convolution of single-item Count sketches can be computed in $O(1)$ time, with respect to the sketch size $m$.} This means that the sketch of a stream can be updated in constant time for each arriving tuple, in contrast to the $O(m)$ time required by the AMS sketch with the Hadamard product. As we will show empirically in Section~\ref{sec:exp-timing}, this translates to sketch updates that are orders of magnitude faster.

\revised{
\begin{definition}[Circular cross-correlation]
\label{def:circ-cross}
The circular cross-correlation $\vx \star \vy$ of any two vectors $\vx, \vy \in \comps^m$ is a vector with elements given by $(\vx \star \vy)_j = \sum_{i=0}^{m-1} \overline{\evx}_i \evy_{(j+i) \bmod m}$ for all $j \in \sqbr{m}$, where $\overline{\evx}_i$ denotes the complex conjugate of $\evx_i$.
\end{definition}
}

\begin{figure}[t]
    \centering
    \begin{tikzpicture}[every node/.style={minimum size=0.5cm-\pgflinewidth, outer sep=2pt}]

\node [draw=none,anchor=east] at (0.1,.52) {$\mPi^{(1)}_{\cdot, a} =$};
\node [draw=none,anchor=east] at (0.1,-0.48) {$\mPi^{(2)}_{\cdot, b} =$};

\node[draw, minimum width=2.5cm] at (1.5, .5) {};
\foreach \i in {0,1,2,...,3} {
    \draw (0.75+\i*0.5, 0.25) -- (0.75+\i*0.5, .75);
}
\node[] at (.5,.5) {0};
\node[] at (.5+.5,.5) {0};
\node[] at (.5+1,.5) {-1};
\node[] at (.5+1.5,.5) {0};
\node[] at (.5+2,.5) {0};

\node [draw=none,anchor=south] at (1.5,-.31) {$\circ$};

\node[draw, minimum width=2.5cm] at (1.5, -.5) {};
\foreach \i in {0,1,2,...,3} {
    \draw (0.75+\i*0.5, -0.25) -- (0.75+\i*0.5, -.75);
}
\node[] at (.5,-0.5) {0};
\node[] at (.5+.5,-0.5) {0};
\node[] at (.5+1,-0.5) {0};
\node[] at (.5+1.5,-0.5) {+1};
\node[] at (.5+2,-0.5) {0};

\node [draw=none,anchor=south] at (1.5,-1.33) {$=$};

\node[draw, minimum width=2.5cm] at (1.5, -1.5) {};
\foreach \i in {0,1,2,...,3} {
    \draw (0.75+\i*0.5, -1.25) -- (0.75+\i*0.5, -1.75);
}
\node[] at (.5,-1.5) {0};
\node[] at (.5+.5,-1.5) {0};
\node[] at (.5+1,-1.5) {0};
\node[] at (.5+1.5,-1.5) {0};
\node[] at (.5+2,-1.5) {0};

\node[draw, minimum width=2.5cm] at (5, .5) {};
\foreach \i in {0,1,2,...,3} {
    \draw (4.25+\i*0.5, .25) -- (4.25+\i*0.5, .75);
}
\node[] at (4,.5) {0};
\node[] at (4+.5,.5) {0};
\node[] at (4+1,.5) {-1};
\node[] at (4+1.5,.5) {0};
\node[] at (4+2,.5) {0};

\node [draw=none,anchor=south] at (5,-.31) {$*$};

\node[draw, minimum width=2.5cm] at (5, -.5) {};
\foreach \i in {0,1,2,...,3} {
    \draw (4.25+\i*0.5, -.25) -- (4.25+\i*0.5, -.75);
}
\node[] at (4,-0.5) {0};
\node[] at (4+.5,-0.5) {0};
\node[] at (4+1,-0.5) {0};
\node[] at (4+1.5,-0.5) {+1};
\node[] at (4+2,-0.5) {0};

\node [draw=none,anchor=south] at (5,-1.33) {$=$};

\node[draw, minimum width=2.5cm] at (5, -1.5) {};
\foreach \i in {0,1,2,...,3} {
    \draw (4.25+\i*0.5, -1.25) -- (4.25+\i*0.5, -1.75);
}
\node[] at (4,-1.5) {-1};
\node[] at (4+.5,-1.5) {0};
\node[] at (4+1,-1.5) {0};
\node[] at (4+1.5,-1.5) {0};
\node[] at (4+2,-1.5) {0};

\node [draw=none,anchor=north west,align=left, red] (formula) at (4.2,-1.83) {$(2 + 3) \bmod{5} = 0$};

\draw[-latex,red] ($(formula.west)+(0.1,0)$) arc
[
    start angle=-90,
    end angle=-160,
    x radius=0.3cm,
    y radius =0.4cm
] ;
\end{tikzpicture}
    \caption{Comparison of the Hadamard product (left) and circular convolution (right) on two single-item Count Sketches. The resulting sketch represents the 2-tuple $(a, b)$.}
    \label{fig:hada-vs-conv}
\end{figure}

\subsection{General formulation by example}

In order to facilitate a better understanding of the proposed method, we begin by demonstrating it through an illustrative example query. By showcasing the estimation process, we aim to provide valuable insights into the inner workings of our method. Following this, we formally present the method through its pseudocode. Furthermore, we prove that our method is an unbiased cardinality estimator for the previously defined family of multi-join queries and provide bounds on the estimation error. A general overview of our estimation procedure is provided in Algorithm~\ref{alg:general-sketch}. In the following description, we shall use the example query from Figure~\ref{fig:example-graph-query}. 

\begin{algorithm}
\caption{General estimation procedure}\label{alg:general-sketch}
\begin{algorithmic}[1]
\Require Relations $R_k$ for $k \in \sqbr{r}$, query $Q(R_0, R_1, \dots, R_{r-1})$, and sketch size $m$.
\Ensure Estimate $X$
\State $s \gets \mathrm{SampleSignHashes}(Q, m)$
\State $h \gets \mathrm{SampleBinHashes}(Q, m)$
\For{\textbf{each} relation $R_k$}
\State $\vc_k \gets \mathrm{CreateSketch}(R_k, s, h, Q, m)$
\EndFor
\State $X \gets \mathrm{GetQueryEstimate}(\vc_0, \vc_1, \dots, \vc_{r-1}, Q, m)$
\end{algorithmic}
\end{algorithm}

\paragraph{Initialization} Our method starts by initializing the necessary hash functions and counters. We sample an independent random \textit{sign function} $s_{\set{u,v}} \colon \sqbr{n} \to \set{-1, +1}$, drawn from a family of 4-wise independent hash functions for every join $\set{u,v}\in E$, represented by an edge in Figure~\ref{fig:example-graph-query}. 
Moreover, each graph component is assigned an independent random \textit{bin function} $h_{\Psi(u)} \colon \sqbr{n} \to \sqbr{m}$, drawn from a family of 2-wise independent hash functions. 
Here, $\Psi(u)$ represents the graph component to which attribute $u$ belongs. A graph component comprises a set of attributes connected by joins, forming a subgraph that is not part of any larger connected subgraph. In the example, there are two graph components: $\set{0, 1, 3}$ and $\set{2, 4}$,
identified by the edge colors in Figure~\ref{fig:example-graph-query}.
\revised{
A bin function is shared within a graph component because all the attributes that form a graph component must, by definition, be joined on equal values. Note that the equal values are mapped to the same bin by using the same bin function.
} 
Lastly, for each relation, a zero vector of $m$ counters is initialized.

\paragraph{Sketching} Once the counters and hash functions are initialized, the tuples from each relation stream are processed. 
When a tuple streams in, it is mapped to a single sign and bin, derived from the signs and bins of the joined attributes.
To determine the sign of a tuple, all the signs of the joined attributes are multiplied together. 
For instance, tuple $i$ from relation $R_1$ is hashed as follows: $s_{\set{1,3}}(i_{1}) s_{\set{1,0}}(i_{1}) s_{\set{2, 4}}(i_{2})$, where $i_u$ is the value for attribute $u$, and $\set{u,v}$ denotes the join between attributes $u$ and $v$.
This is because $R_1$ has two joined attributes, and one of those is joined twice.
Formally, the sign of a tuple $i$ from relation $R_k$ is given by:
\begin{align}
    S_k(i) = \prod_{u \in \Omega(R_k)} \prod_{v \in \Gamma(u)} s_{\set{u, v}}(i_u)
\end{align}
To determine the bin of the tuple, the bin indices of all the joined attributes are summed, followed by taking the modulo $m$. Continuing our example, we have: $(h_{\Psi(1)}(i_{1}) + h_{\Psi(2)}(i_{2})) \bmod{m}$ because $R_1$ has two joined attributes. The bin index of a tuple $i$ from relation $R_k$ is formally given by:
\begin{align}
    H_k(i) = \rdbr*{\sum_{u \in \Omega(R_k)} h_{\Psi(u)}(i_u)} \bmod{m}
\end{align}
Subsequently, the sign of the tuple multiplied by the change of frequency is added to the counter at the bin index of the tuple. 

The sketching process for a tuple is equivalent to circular convolution between the sketches for each joined attribute value in the tuple. Since the individual sketches have only one non-zero value, the result of the circular convolution also has one non-zero value, which can be computed in constant time with respect to the sketch size, as explained earlier. The pseudocode for the general sketch creation procedure is provided in Algorithm~\ref{alg:create-sketch}. The sketch $\vc_k$ for relation $R_k$ is formally stated as follows:
\begin{align}
    \evc_{k,j} = \sum_{i \in I_k \colon H_k(i) = j} \etF_k(i) S_k(i) \label{eq:our-sketch}
\end{align}
\revised{where $\etF_k(i)$ denotes the frequency of tuple $i$ in relation $R_k$. The tuple $i$ is from the domain $I_k = \sqbr{n} \times \cdots \times \sqbr{n}$ of relation $R_k$}.

\begin{algorithm}
\caption{Sketch creation procedure}\label{alg:create-sketch}
\begin{algorithmic}[1]
\Function{CreateSketch}{$R_k, s, h, Q, m$}
\State $\vc_k \gets \vzero$ \Comment{Size: $m$}
\For{\textbf{each} tuple $(i, \Delta)$ in $R_k$} \Comment{The stream of tuples}
\State $x = 1$ \Comment{For accumulation of signs}
\State $j = 0$ \Comment{For accumulation of bins}
\For{\textbf{each} attribute $u$ in $\Omega(R_k)$}
\State $j \gets j + h_{\Psi(u)}(i_{u})$
\For{\textbf{each} attribute $v$ in $\Gamma(u)$}
\State $x \gets s_{\set{u, v}}(i_{u}) x$
\EndFor
\EndFor
\State $j \gets j \bmod{m}$
\State $\evc_{k,j} \gets \evc_{k,j} + x\Delta$
\EndFor
\State \Return sketch $\vc_k$
\EndFunction
\end{algorithmic}
\end{algorithm}

\paragraph{Inference} Upon creating the sketches for each relation, we can proceed to estimate the query's cardinality by combining sketches using either the Hadamard product or circular cross-correlation. 
The computation consists of summations over the sketch size for each graph component. The sketch for each relation is indexed by the sums of the graph components that have an attribute in that relation. Sketches of relations with multiple joined attributes will, therefore, also have multiple indices, and these are summed to obtain the final index. The sketch values inside the sums are all multiplied. 
An estimate of the example query is then obtained as follows: $\sum_{j_0=0}^{m-1} \sum_{j_1=0}^{m-1} \evc_{0,j_0}\evc_{2,j_0}\evc_{1,(j_0+j_1)\bmod{m}}\evc_{3,j_1}$ because there are two graph components, and $R_1$ is part of both. This can be factorized as the Hadamard product between sketches $\vc_0$ and $\vc_2$ whose result is circular cross-correlated with $\vc_1$, followed by another Hadamard product with $\vc_3$, and lastly a summation over the elements. 
A cardinality estimate $X$ is formally obtained as follows: 
\begin{align}
    X = \sum_{j \in J} \prod_{k=0}^{r-1} \evc_{k, G_k(j)} \label{eq:our-est}, \quad \text{with} \quad G_k(j) = \rdbr*{\sum_{u \in \Omega(R_k)} j_{\Psi(u)}} \bmod{m}
\end{align}
where $J$ is the cross product of bin domains $\sqbr{m} \times \cdots \times \sqbr{m}$ for each graph component.

Computing the estimates naively using Equation~\ref{eq:our-est} has an exponential time complexity with the number of graph components. However, this can be improved significantly by factorizing the problem. We can then rely on the fact that circular cross-correlation can be computed efficiently in $O(m \log m)$ time using the fast Fourier transform (FFT).
To perform this efficient estimation process methodically, one starts by selecting any joined attribute, say attribute 4 in our example. The process now aggregates all the sketches towards attribute 4 to obtain the estimate. This is implemented as a depth first traversal of the join graph with attribute 4 as the root node of a rooted tree and attributes 0 and 3 as the leaves. 
The general procedure for combining sketches to efficiently compute an estimate of the query is provided in Algorithm~\ref{alg:estimate-query}. 
In the pseudocode, we ensure that the recursion only moves away from any selected root attribute $o \in \sqbr{w}$ by keeping track of the visited nodes $V$. The functions (I)FFT denote the (inverse) fast Fourier transform. This procedure reduces the inference time complexity to $O(rm\log{m})$, that is, nearly linear with respect to the sketch size.

\begin{algorithm}
\caption{Estimation procedure}\label{alg:estimate-query}
\begin{algorithmic}[1]
\Function{GetQueryEstimate}{$\vc_0, \vc_1, \dots, \vc_{r-1}, Q, m$}
\State $o \gets \mathrm{AnyJoinedAttribute}(Q)$ \Comment{The root attribute}
\State $V \gets \set{}$ \Comment{Global set of visited attributes} 
\State $X \gets \mathrm{SumElements}(\mathrm{CombineSketches}(o, V, m))$
\State \Return estimate $X$
\EndFunction

\Function{CombineSketches}{$u, V, m$}
\State $R_k \gets \mathrm{RelationOf}(u)$
\State $\vx \gets \vc_k$ \Comment{Sketch of relation $R_k$}
\State $\mathrm{add}(V, u)$ \Comment{Adds attribute $u$ to the visited set}
\LineComment{Recurse through the other attributes in the relation.}
\For{\textbf{each} attribute $u'$ in $\Omega(R_k) \setminus \set{u}$}
\State $\mathrm{add}(V, u')$
\State $\va \gets \vone$ \Comment{Size: $m$}
\LineComment{By the definition of $\Omega$ there is at least one iteration.}
\For{\textbf{each} attribute $v$ in $\Gamma(u')$}
\State $\va \gets \mathrm{CombineSketches}(v,V, m) \circ \va$
\EndFor
\LineComment{Efficient circular cross-correlation}
\State $\vx \gets \mathrm{IFFT}(\overline{\mathrm{FFT}(\va)} \circ \mathrm{FFT}(\vx))$
\EndFor
\LineComment{Recurse over the attributes joined with the current.}
\For{\textbf{each} attribute $v$ in $\Gamma(u) \setminus V$}
\State $\vx \gets \mathrm{CombineSketches}(v,V, m) \circ \vx$
\EndFor
\State \Return intermediate sketch $\vx$
\EndFunction
\end{algorithmic}
\end{algorithm}

\subsection{Analysis}
\label{sec:analysis}

Now that we have discussed the estimation procedure, we present a theoretical analysis of the proposed method. We show that it is an unbiased estimator for the cardinality of multi-join queries in Theorem~\ref{thm:ours}, and provide guarantees on the estimation error. Lastly, we provide the time complexity for each estimation stage.

\begin{theorem}
\label{thm:ours}
Given an acyclic query of relations $R_k$ for $k \in \sqbr{r}$, let Equation~\ref{eq:our-sketch} provide the sketch for each relation and Equation~\ref{eq:our-est} the cardinality estimate $X$ of the query, then we have: 
\begin{align*}
    \E\sqbr{X} &= \sum_{i\in I_0 \times \cdots \times I_{r-1}}  \etF_{0}(i) \cdots \etF_{r-1}(i) \prod_{\set{u,v} \in E} \1(i_u = i_v)\\
    \Var(X) &\leq \frac{1}{m}3^{r-1} \prod_{k=0}^{r-1} \norm*{\tF_k}^2_2
\end{align*}
\end{theorem}

\begin{proof}
We present the proof in Appendix~\ref{sec:proofs}.
\end{proof}

Using the Chebyshev inequality and the upper bound on the variance from Theorem~\ref{thm:ours}, we can bound the absolute estimation error by $\epsilon > 0$ with $m \geq 3^{r} \epsilon^{-2} \prod_{k=0}^{r-1} \norm*{\tF_k}^2_2$. Furthermore, to guarantee the error with probability at most $1 - \delta$, one selects the median of $l = O(\log{1/\delta})$ i.i.d. estimates by the Chernoff bound~\citep{alon1996space}. The exponential term $3^{r}$ indicates that accurately estimating queries involving many relations quickly becomes infeasible. \revised{It remains an open problem whether this exponential dependence can be improved.} However, as we will show in the following section, for moderate sized queries, involving up to 6 relations, our estimation accuracy constitutes a significant improvement over the baselines.

\begin{table}[h]
    \centering
    \caption{Comparison of time complexity by stage}
    \label{tab:time-complexity}
    \begin{tabular}{l|ccc}
    \toprule
    Method & Initialization & Update & Inference \\
     \midrule
    AMS & $O(rlm)$ & $O(rlm)$ & $O(rlm)$ \\
    COMPASS (partition) & $O(lm^r)$ & $O(rl)$ & $O(lm^r)$ \\
    COMPASS (merge) & $O(rlm)$ & $O(rl)$ & $O(rlm^r)$ \\
    \textit{Ours} & $O(rlm)$ & $O(rl)$ & $O(rlm\log{m})$\\
    \bottomrule
    \end{tabular}
\end{table}

In Table~\ref{tab:time-complexity}, we present the time complexity of each estimation stage, comparing our method with the AMS-based technique by \citet{dobra2002processing} and the two variations of COMPASS (partition and merge) \cite{izenov2021compass}.
The symbols $r$, $l$, and $m$ denote the number of relations, medians, and the sketch size, respectively. 
The update time of our method for each incoming tuple is remarkably efficient, with a time complexity of only $O(r \log{1/\delta})$. 
The efficient update time complexity, independent of the estimation error $\epsilon$, enables the sketching of high-throughput streams even when requiring a high level of accuracy.
While COMPASS also has fast updates, its exponential dependence on $r$ during inference limits its practical use even for moderately sized queries.
Our method achieves fast updates yet introduces only an additional $\log m$ term during the inference stage, compared to the AMS baseline. 
This slight increase in inference time is negligible when considering the substantial improvement in update time. 
For instance, our experiments go up to $m=10^{6}$ with $l=5$, which means that our method achieves roughly $10^{6}$ times faster sketch updates, while having only $\log_2(10^{6}) \approx 20$ times slower inference.
As a result, our method effectively minimizes the overall estimation time in various crucial scenarios.
These claims are further supported by our empirical results, which are detailed in Section~\ref{sec:experiments}.

\revised{
\subsection{Integration with query optimizers}
\label{sec:integrate-query-optimizer}

The quintessential application of our proposed method is the cardinality estimator within query optimizers. 
Query optimizers use a plan enumeration algorithm to find a good join order. 
The cost of a join order dependents upon the sizes of the intermediate results. 
The cardinality estimator's role is to provide the estimates for these intermediate sizes~\citep{lan2021survey}. 
Each intermediate cardinality can be expressed as a sub-query which our method can estimate. 
The sketches for all the evaluated sub-queries can be created in a single pass over the data. 
Typically, each sub-query requires its own sketches; however, in cases where an attribute is joined multiple times, the sketches can be reused for each join involving that attribute. For example, to decide the join order of joins $\set{0,1}$ and $\set{1, 3}$ in Figure~\ref{fig:example-graph-query}, the sketch for attribute~1 can be reused for attributes 0 and 3.
In Section~\ref{sec:postgresql}, we demonstrate the improvement in query execution time after integrating our proposed cardinality estimator into the query optimizer of PostgreSQL.
}

\section{Experiments}
\label{sec:experiments}

In this section, we conduct an empirical analysis to evaluate the effectiveness of our estimator and compare it to various baseline approaches. These baselines include the AMS-based method proposed by \citet{dobra2002processing} and the two variations of COMPASS~\citep{izenov2021compass}, namely partition and merge. Our primary objective is to assess the accuracy of our estimator against the baselines for a specified memory budget. Furthermore, we compare the initialization, sketching, and inference times of our method with those of the baselines. Secondly, we compare the estimation error \revised{and execution time} of our method with the four data-driven machine learning techniques discussed in Section~\ref{sec:other}: DeepDB~\citep{hilprecht13deepdb}, BayesCard~\citep{wu2020bayescard}, NeuroCard~\citep{yang2020neurocard}, and FLAT~\citep{zhu2021flat}. 
\revised{Finally, we evaluate the impact of our cardinality estimator on the query execution time of PostgreSQL.}

We implemented both the sketching baselines and our method using the PyTorch tensor library~\citep{paszke2019pytorch}.
The hash functions are implemented using efficient random polynomials over a Mersenne prime field~\citep{ahle2020power}.
All experiments were conducted on an internal cluster of Intel Xeon Gold 6148 CPUs. Each experiment utilized a single CPU and 24~GB of memory.
The source code for the experiments, extended results, and cardinality estimates are available online\footnote{\label{link:github}Source code: \url{https://github.com/mikeheddes/fast-multi-join-sketch}}.

\begin{figure}
  \centering
  \subimport*{figures/}{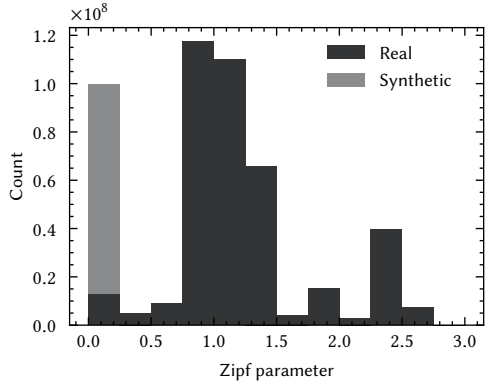}
  \caption{Total number of entries across all columns of both the STATS and IMDB databases, grouped by the best fit Zipf parameter of each column. Synthetic entries refer to the id and md5sum columns, which are unique by design, the real entries include all other columns.}
  \label{fig:zipf-param-dist}
\end{figure}

\subsection{Databases and queries}
\label{sec:databases}

The limitations of synthetic databases in accurately reflecting real-world performance have been widely acknowledged in the literature~\citep{leis2015good, han2021cardinality}. To address this concern, our experiments were conducted using two established benchmarking databases containing real data: the IMDB database~\citep{leis2015good}, which encompasses information on movies, actors, and their associated production companies, and the STATS database~\cite{han2021cardinality}, which comprises user-contributed content from the Stats Stack Exchange network. 
To provide insights into the characteristics of the databases, Table~\ref{tab:database-stats} presents statistics detailing their sizes. Additionally, Figure~\ref{fig:zipf-param-dist} showcases the distribution skewness of the database entries, highlighting the significant skewness often observed in real data~\citep{dobra2002processing, leis2015good, manerikar2009frequent,yang2017pyramid,roy2016augmented}.
Our experimentation covers all the 146 STATS-CEB and 70 JOB-light queries, in addition to the 3299 associated sub-queries from the cardinality estimation benchmark~\citep{han2021cardinality}. These queries collectively represent a diverse range of real-world workloads. 

\begin{table}[h]
    \centering
    \caption{Database size statistics}
    \label{tab:database-stats}
    \begin{tabular}{l|ccc}
        \toprule
         Database & Relations & Tuples & Storage size \\
         \midrule
         IMDB & 21 & 74.2M & 3.88~GB \\
         STATS & 8 & 1.03M & 39.6~MB \\
         \bottomrule
    \end{tabular}
\end{table}

The key feature of our proposed method is its ability to efficiently estimate multi-join queries. As previously mentioned, our motivation for this capability is rooted in the prevalence of such queries in real-world scenarios. To further substantiate this motivation, we conducted an analysis of all the queries in both the cardinality estimation benchmark~\citep{han2021cardinality} and the join order benchmark~\citep{leis2015good}. The results, displayed in Table~\ref{tab:max-joined-attributes}, indicate that 44\% of the relations in the queries are involved in multiple joins, with single joins being the most common at 57\%. Notably, 97\% of the queries contain at least one relation which participates in multiple joins. These statistics underscore the significance of supporting multi-join queries to effectively address the majority of real-world query scenarios.

\begin{table}[h]
    \centering
    \caption{Percentage of relations among all queries by their number of joins, and the percentage of queries by their relation with the maximum number of joins.}
    \label{tab:max-joined-attributes}
    \begin{tabular}{l|ccccc}
    \toprule
        Joins & 1 & 2 & 3 & 4 & 5+ \\
    \midrule
        Relations & 57\% & 12\% & 9\% & 9\% & 12\%\\
        Queries & 3\% & 24\% & 30\% & 20\% & 23\%\\
    \bottomrule
    \end{tabular}
\end{table}

Following \citet{izenov2021compass}, in the experiments the filter predicates of each query are processed during ingestion of the tuples from their respective relation streams.  
In many streaming algorithms, the query is assumed to be known in advance of the stream. Consequently, filtering the tuples at the time of ingestion offers advantages in terms of performance and accuracy. This approach eliminates the need to update the sketch for tuples that do not satisfy the filters. In addition, the estimation accuracy is significantly improved as some data is already filtered out from the sketches \citep{izenov2021compass}.
However, it is worth mentioning that sketching methods, including the one presented, are also capable of handling filter predicates during inference. This can be achieved by treating the filters as joins with imaginary tables, a technique employed, for example, by \citet{cormode2011sketch} and \citet{vengerov2015join}.
Lastly, in our experiments, we report the median of $l=5$ i.i.d. estimates as the cardinality estimate for all sketching methods.

\subsection{Estimation accuracy}
\label{sec:exp-accuracy}

\begin{figure*}[ht]
  \centering
  \subimport*{figures/}{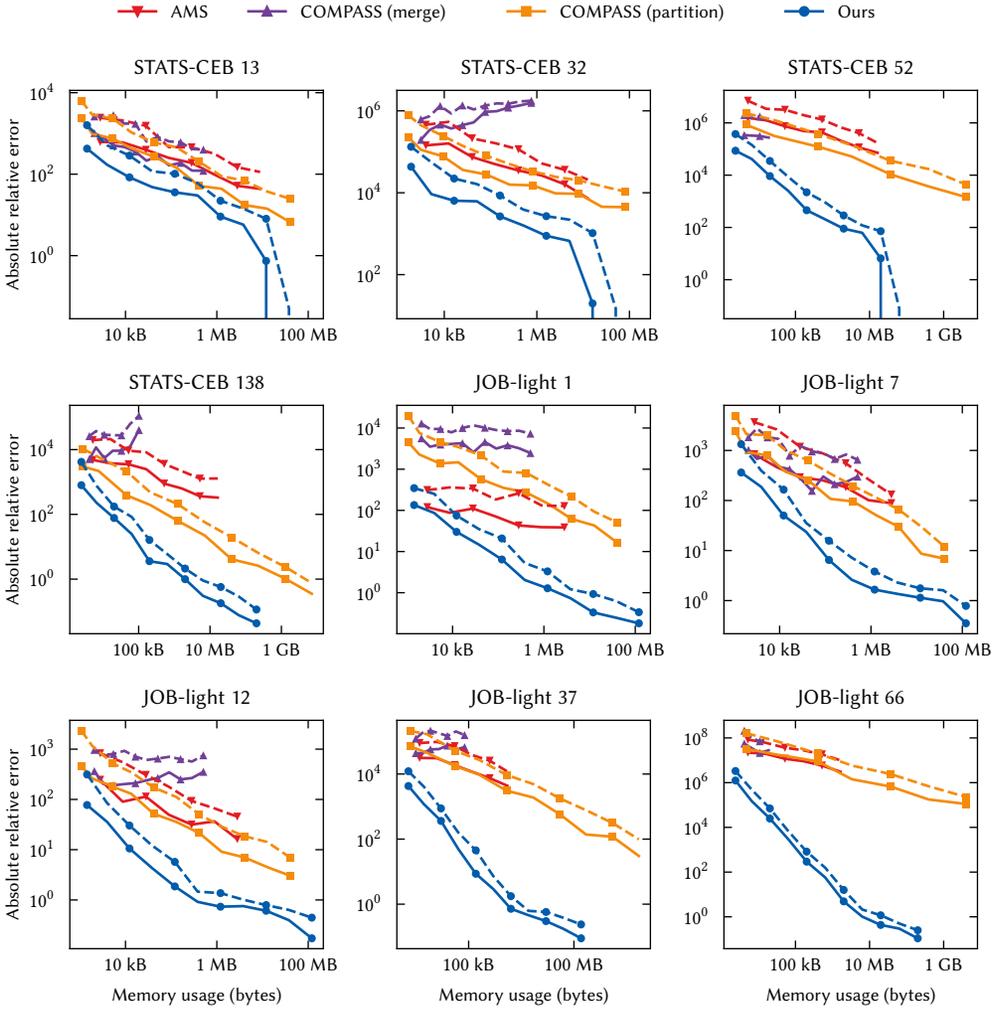} 
  \caption{Absolute relative error of our method compared to the baselines at varying sketch sizes. Solid lines represent the median error and dashed lines denote the 95th percentile. The memory usage includes the counters of the sketches and the random seeds for the hash functions, but excludes the space needed for the intermediate inference calculations. }
  \label{fig:error}
\end{figure*}

We first compare the estimation accuracy of the different sketching methods in terms of the absolute relative error, defined as $\abs{y - \hat{y}}$ divided by $\max(y, 1)$, where $y$ is the true cardinality and $\hat{y}$ its estimate. This metric aligns with the formulation of the theoretical error bound outlined in Section~\ref{sec:analysis}. 
In Figure~\ref{fig:error}, we present the median and the 95th percentile of the error at varying sketch sizes.
The statistics are derived from 30 repetitions, each with distinct random initializations. 
The results are presented for a representative subset of the queries, necessitated by space constraints, but detailed results for all 216 queries can be found online\footref{link:github}.

It is important to note that the experiments for AMS and COMPASS (merge) do not extend to the highest memory usage levels. 
This limitation arises due to the extensive time required to run the AMS-based experiments, which increases exponentially with each additional data point, rendering their execution quickly infeasible. Additionally, COMPASS (merge) encountered memory constraints during the inference stage, as its memory demand grows exponentially with the number of joins. Notice that the depicted memory usage excludes intermediate representations during inference, thus understating the actual memory required for COMPASS (merge).

Upon analysing the results, we observe that the proposed method delivers comparable or lower error rates across all queries, often demonstrating orders of magnitude greater accuracy. 
\revised{
The proposed method achieves zero error on many queries with large sketches. In realistic sketching applications, a margin of error is acceptable; therefore, sketches ranging from 1 to 10~MB could be employed for the STATS-CEB and JOB-light queries.
}
For certain queries, like JOB-light 37 and 66, our method not only achieves significantly lower error but also demonstrates a more rapid reduction in error with each increase in memory.

\begin{figure}[h]
  \centering
  \subimport*{figures/}{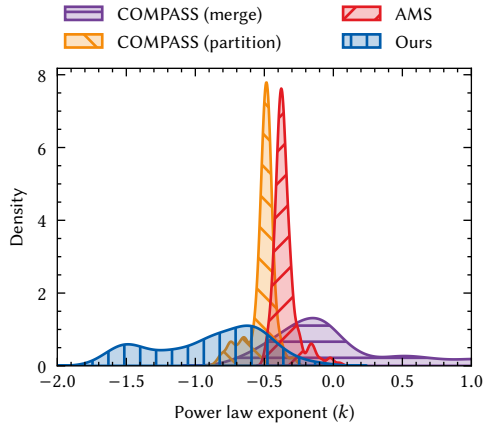}
  \caption{Kernel density estimates of the power law exponents from the absolute relative error plots for all 216 queries. A higher concentration on the left signifies a greater improvement in accuracy as the memory budget increases.}
  \label{fig:error-slope}
\end{figure}

To assess the rate at which our method improves the estimation error compared to the baselines, Figure~\ref{fig:error-slope} presents kernel density estimates of the slopes derived from the absolute relative error results for all 216 queries. These slopes are obtained by least-squares linear regression of the data for each method and query in log-log space. This means that the slope represents the exponent of a power law relationship, where the error at memory usage $m$ is given by $am^k$, with $k$ as the slope and $a$ as the error at $m=1$. 

Remarkably, our method exhibits a significantly faster reduction in error, highlighting its ability to achieve high accuracy with substantially less memory compared to the baselines. 
Considering that the real data in our experiments is skewed, as indicated in Figure~\ref{fig:zipf-param-dist}, 
we speculate that our method successfully inherits and expands upon the advantages associated with the Count sketch, particularly its effectiveness in handling skewed data. In the context of multi-joins, our method capitalizes on these benefits, demonstrating its ability to compute accurate cardinality estimates.

\subsection{Execution times}
\label{sec:exp-timing}

\begin{figure*}[ht]
  \centering
  \resizebox{\textwidth}{!}{\subimport*{figures/}{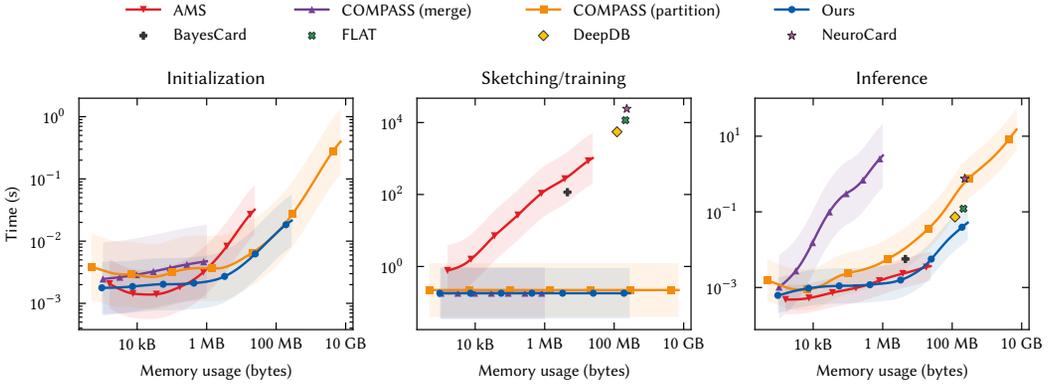}}
  \caption{
  Execution times for the three stages (initialization, sketching/training, and inference) of the baselines and our method at varying sketch/model sizes. The plots show the best fit of a Gaussian process regression of the data in log-log space, along with standard deviation. The memory usage includes the counters of the sketches and the random seeds for the hash functions. \revised{The data for BayesCard~\citep{wu2020bayescard}, FLAT~\citep{zhu2021flat}, DeepDB~\citep{hilprecht13deepdb}, and NeuroCard~\citep{yang2020neurocard} is obtained from \citet{han2021cardinality}.} }
  \label{fig:timing}
\end{figure*}

In the second set of experiments, we look into the execution time of our method and how it compares to the baselines across varying sketch sizes. In Figure~\ref{fig:timing}, we present the execution times for each stage of the estimation process: initialization, sketching, and inference. The figure shows the best fit of a Gaussian process regression to the experimental results from all queries, totaling $232{,}323$ experiments.
\revised{It also contains data for the learning-based methods which will be discussed in the following section.}
The individual timing results for all queries are provided online\footref{link:github}.

While the initialization time exhibits a similar trend for all methods, in sketching time there is a notable disparity between AMS and the other methods. This disparity directly reflects the difference in update time complexity.
\revised{
The methods also differ in initialization time complexity, but this is not visible in the timing results because they are plotted with respect to their memory usage rather than the sketch size. That is, for a given sketch size COMPASS (partition) allocates more memory, but for a given memory budget, all methods have similar initialization time.
}
Among the inference times, our method demonstrates remarkable overall efficiency. Even for the most complex queries with the largest sketch size, our method computes its estimate within ten seconds. In contrast, the AMS-based method requires hours to compute estimates, with the majority of that time spent on sketching, all while delivering higher error rates.

To further validate the fast update time of our method, we assessed the maximum stream throughput for all baselines, as outlined in Table~\ref{tab:throughput}. The memory usage includes the counters of the sketches and the random seeds for the hash functions. The results were obtained by performing linear least-squares regression on the throughput measurements for each method on all queries.
For smaller sketch sizes, the AMS-based method can achieve a throughput similar to the Count sketch-based approaches. 
However, when working with larger sketch sizes required for high estimation accuracy, the AMS-based method becomes limited to handling just a few hundred tuples per second. This significant limitation severely restricts the practical usability of AMS in streaming scenarios.

\begin{table}[h]
    \centering
    \caption{Throughput in tuples processed per second}
    \label{tab:throughput}
    \setlength{\tabcolsep}{4pt} %
    \begin{tabular}{l|cccccc}
    \toprule
        Memory usage & 1~kB & 10~kB & 100~kB & 1~MB & 10~MB\\
    \midrule
        AMS & 5.2M & 576k & 63.5k & 7.0k & 774\\
        COMPASS (partition) & 5.9M & 5.9M & 5.9M & 5.9M & 5.9M\\
        COMPASS (merge) & 6.3M & 6.2M & 6.0M & 5.8M & 5.6M\\
        \textit{Ours} & 7.0M & 6.8M & 6.6M & 6.5M & 6.3M\\
    \bottomrule
    \end{tabular}
\end{table}

\subsection{Comparison with learning-based methods}
\label{sec:q-error}

\begin{figure}[h]
  \centering
  \subimport*{figures/}{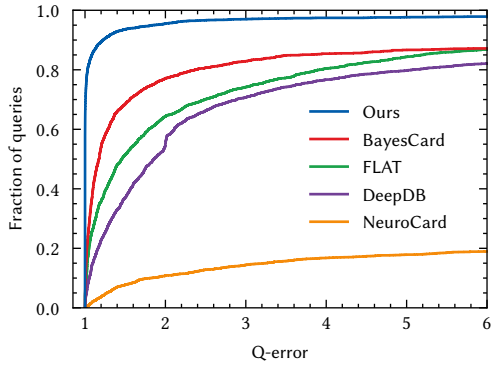}
  \caption{Cumulative distribution function of the q-error for the STATS-CEB and JOB-light sub-queries. The legend follows the ordering of the lines in the figure.}
  \label{fig:sub-queries-error-cdf}
\end{figure}

In this set of experiments, we compare the cardinality estimation performance of our proposed method with the four data-driven machine learning techniques discussed in Section~\ref{sec:other}. 
\revised{Specifically, we compare their execution time and estimation quality.} 
To evaluate the quality of cardinality estimates, we employ the q-error metric, defined as $\max(y / \hat{y}, \hat{y} / y)$ if $\hat{y} > 0$ and $\infty$ otherwise~\cite{moerkotte2009preventing}.
Figure~\ref{fig:sub-queries-error-cdf} presents the cumulative distribution function of the q-error for all 3299 sub-queries, showing the fraction of queries that were estimated within a certain q-error. Our method was configured with $m=1{,}000{,}000$ bins, resulting in an average estimation time of 0.30 seconds and consuming an average of 137~MB of memory. To maintain consistency with the results of the learning methods, as obtained from the cardinality estimation benchmark~\citep{han2021cardinality}, our method estimated the cardinality of each sub-query only once. 
We provide our cardinality estimates for the sub-queries together with the source code\footref{link:github} to facilitate further comparisons with the proposed method and to ensure reproducibility.

The results presented in Figure~\ref{fig:sub-queries-error-cdf} highlight the superior performance of the proposed method, which delivers error-free estimates for approximately 70\% of the sub-queries.
Furthermore, our method achieves q-error values of less than 2 for about 95\% of the sub-queries. In contrast, even the best-performing learning-based method, BayesCard, achieves this level of accuracy for less than 80\% of the sub-queries. 
These findings demonstrate the remarkable estimation accuracy of our proposed estimator. Moreover, our sketching approach provides theoretical guarantees on the estimation error which are lacking for the learning-based methods.

\revised{
Our proposed method is also notably efficient when compared to the learning-based methods.
As depicted in Figure~\ref{fig:timing}, the training phase of the learning-based methods requires 3 to 5 orders of magnitude more time than creating our sketches. 
In addition, Table~\ref{tab:throughput} shows that our method can handle over 6 million updates per second. This is because our method simply adds another item to the sketch. BayesCard, on the other hand, takes 12 seconds to process an update \citep{han2021cardinality}, and the other learning-based methods take several minutes. Consequently, these learning-based methods prove impractical for scenarios involving frequent updates.
}

\revised{
\subsection{PostgreSQL query execution time}
\label{sec:postgresql}

In the last set of experiments, we evaluate the impact of our cardinality estimator on the query execution time of PostgreSQL. 
We utilized the evaluation setup devised by \citet{han2021cardinality}, they modified PostgreSQL to enable the injection of cardinality estimates for all the sub-queries of the STATS-CEB and JOB-light queries. 
We compare our method against PostgreSQL's own cardinality estimator as well as the aforementioned learning-based methods. 
Table~\ref{tab:psql-exec} shows the total execution time for all the queries. The injected sub-query cardinality estimates are those reported in Section~\ref{sec:q-error}. In the results, PostgreSQL refers to the default PostgreSQL cardinality estimator, and True Cardinality denotes an oracle method with access to the actual intermediate sizes. 

Our proposed method achieved the lowest total execution time with an improvement of 43\% compared to PostgreSQL. On STATS-CEB, our method improves over PostgreSQL by 48\%, falling just short of FLAT, which showed an improvement of 55\%. On JOB-light, our method achieved equivalent execution time to the True Cardinality, while most other learning-based methods, with the exception of BayesCard, do not improve over PostgreSQL. These results underscore the practical advancement enabled by our proposed method due to its superior estimation accuracy, in addition to its exceptional efficiency.

\begin{table}[h]
    \centering
    \caption{PostgreSQL plan execution time and percentage improvement using different cardinality estimators.}
    \label{tab:psql-exec}
    \setlength{\tabcolsep}{4pt} %
    \begin{tabular}{l|cccccc}
    \toprule
         Method & \multicolumn{2}{c}{STATS-CEB} & \multicolumn{2}{c}{JOB-light} & \multicolumn{2}{c}{Total} \\
     \midrule
         PostgreSQL & 4.04~h & 0\% & 1.08~h & 0\% & 5.13~h & 0\% \\ 
         True Cardinality & 1.78~h & 56\% & 0.84~h & 23\% & 2.62~h & 49\% \\
         BayesCard & 2.42~h & 40\% & 0.87~h & 20\% & 3.29~h & 36\% \\
         FLAT & \textbf{1.82~h} & \textbf{55\%} & 1.73~h & -60\% & 3.55~h & 31\% \\
         DeepDB & 2.24~h & 45\% & 1.81~h & -67\% & 4.05~h & 21\% \\
         NeuroCard & 4.55~h & -13\% & 2.51~h & -132\% & 7.06~h & -38\% \\
        \textit{Ours} & 2.09~h & 48\% & \textbf{0.83~h} & \textbf{23\%} & \textbf{2.92~h} & \textbf{43\%} \\
     \bottomrule
    \end{tabular}
\end{table}

}

\section{Conclusion}
\label{sec:conclusion}

We have introduced a new sketching method that significantly enhances the cardinality estimation for multi-join queries. The proposed approach provides fast update times, which remain constant irrespective of the required estimation accuracy. This crucial feature allows for efficient processing of high-throughput streams, all the while delivering superior estimation accuracy compared to state-of-the-art baseline approaches. This is substantiated by our bound on the estimation error, as well as our empirical findings.
Our results underscore the practical suitability of the proposed method for applications such as query optimization and approximate query processing, surpassing the capabilities of previous methods.
The presented method successfully addresses the longstanding challenge of integrating the key advantages of the Count sketch with the AMS-based method for multi-join queries.

\appendix
\section{Mean and variance}
\label{sec:proofs}

In this appendix, we present the proofs for Theorems \ref{thm:dobra} and \ref{thm:ours}.

\begin{proof}[Proof for Theorem~\ref{thm:dobra}]
By the definitions of $X$ and $\vc_{k,j}$, with $I = I_0 \times \cdots \times I_{r-1}$, and using the linearity of expectation, we get:
\begin{align*}
\E\sqbr{X} = \frac{1}{m} \sum_{j=0}^{m-1} \sum_{i \in I} \E\sqbr*{\prod_{k=0}^{r-1} \etF_{k}(i) \prod_{u \in \Omega(R_k)} \prod_{v \in \Gamma(u)} s_{j, \set{u, v}}\rdbr{i_u}}
\end{align*}
By construction of the sketches, the sign functions are independent across different joins and there are exactly two occurrences of each sign function, one for each end of a join edge. We can thus write:
\begin{align*}
\E\sqbr{X} = \frac{1}{m} \sum_{j=0}^{m-1} \sum_{i \in I} \rdbr*{\prod_{k=0}^{r-1}\etF_{k}(i)} \prod_{\set{u,v} \in E} \E\sqbr*{s_{j,\set{u,v}}(i_u)s_{j,\set{u,v}}(i_v)}
\end{align*}
Since $\E\sqbr{s(a)s(b)} = \1(a=b)$, we get the desired expectation.

For the variance, all the dimensions of the sketches are i.i.d., and since $\Var(X) = \E[X^2] - \E[X]^2$, for any $j \in \sqbr{m}$, we have:
\begin{align*}
    \Var(X) = \frac{1}{m} \Var\rdbr*{\prod_{k=0}^{r-1} \evc_{k,j}} \leq \frac{1}{m}\E\sqbr*{\rdbr*{\prod_{k=0}^{r-1} \evc_{k,j}}^2}
\end{align*}
By the definition of $\vc_{k,j}$ and the linearity of expectation, we have:
\begin{align*}
    \Var(X) \leq \frac{1}{m}\sum_{i\in I} \sum_{i'\in I}  \etF_{0}(i) \cdots \etF_{r-1}(i)   \etF_{0}(i') \cdots \etF_{r-1}(i') \E\sqbr*{\prod_{k=0}^{r-1} \prod_{u \in \Omega(R_k)} \prod_{v \in \Gamma(u)} s_{j, \set{u, v}}\rdbr{i_u} s_{j, \set{u, v}}\rdbr{i'_u}}
\end{align*}
Taking into account that each sign function occurs twice (now four times since the value is squared), we obtain:
\begin{align*}
    \Var(X) \leq{}&\frac{1}{m}\sum_{i\in I} \sum_{i'\in I}  \etF_{0}(i) \cdots \etF_{r-1}(i)   \etF_{0}(i') \cdots \etF_{r-1}(i') \\&\prod_{\set{u,v} \in E} \E\sqbr*{s_{j, \set{u, v}}\rdbr{i_u} s_{j, \set{u, v}}\rdbr{i'_u}s_{j, \set{u, v}}\rdbr{i_v} s_{j, \set{u, v}}\rdbr{i'_v}}
\end{align*}
By the four-wise independence of the sign functions, the expected value is one if there are two equal pairs or if all values are equal, and zero otherwise. Therefore, we have that:
\begin{align*}
    &\E\sqbr*{s_{j, \set{u, v}}\rdbr{i_u} s_{j, \set{u, v}}\rdbr{i'_u}s_{j, \set{u, v}}\rdbr{i_v} s_{j, \set{u, v}}\rdbr{i'_v}} \\ &= \1\rdbr*{(i_u = i_v \land i'_u = i'_v) \lor (i_u = i'_u \neq i_v = i'_v) \lor (i_u = i'_v \neq i_v = i'_u)}
\end{align*}
For each join, there are three disjunctions which are conjoined over all the joins. By the distributivity property of conjunction over disjunction, there are thus $3^{\abs{E}}$ disjunctions in total. Since the queries are acyclic, $\abs{E} = r-1$. The variance is bound by the sum over all disjunctions, where intersections are counted double. Using the Cauchy–Schwarz inequality, each disjunction itself is bound by the product of squared frequency norms. This gives the desired upper bound on the variance.
\end{proof}

\begin{proof}[Proof for Theorem~\ref{thm:ours}]
By the definitions of $X$ and $S_k$, with $I = I_0 \times \cdots \times I_{r-1}$, and the linearity of expectation, we have:
\begin{align*}
    \E\sqbr{X} = \sum_{j\in J} \sum_{i \in I} \etF_{0}(i) \cdots \etF_{r-1}(i) \E\sqbr*{\prod_{k=0}^{r-1} \1\rdbr*{H_k(i) = G_k(j)} \prod_{u \in \Omega(R_k)} \prod_{v \in \Gamma(u)} s_{\set{u, v}}(i_u)}
\end{align*}
By the definitions of $H_k$ and $G_k$, since the sign and bin functions are independent, and using again the observation that each sign function occurs twice, we can write:
\begin{align*}
    \E\sqbr{X} ={}&\sum_{i \in I} \etF_{0}(i)  \cdots \etF_{r-1}(i) \rdbr*{\prod_{\set{u,v} \in E} \1(i_u = i_v)}\\& \sum_{j\in J} \E\sqbr*{\prod_{k=0}^{r-1} \1\rdbr*{\sum_{u \in \Omega(R_k)} h_{\Psi(u)}(i_u) - j_{\Psi(u)} \equiv 0 \pmod{m}}}
\end{align*}
By isolating the case where all $i_u = i_v$, all the attribute values in the same graph component must be equal.
Since each independent bin function is thus called with only one distinct value, the expected value of the bin functions is $m^{-(w-r+1)}$, which is exactly the reciprocal of $\abs{J}$, giving the desired expectation.

For the variance, we have that $\Var(X) = \E\sqbr{X^2} - \E\sqbr{X}^2$, where:
\begin{align*}
    \E\sqbr{X}^2 ={}& \sum_{i \in I}\sum_{i' \in I} \rdbr*{\prod_{k=0}^{r-1} \etF_k(i)\etF_k(i')} \prod_{\set{u,v} \in E} \1\rdbr{i_u=i_v\land i'_u = i'_v}\\
    \E\sqbr{X^2} ={}& \sum_{i \in I}\sum_{i' \in I} \rdbr*{\prod_{k=0}^{r-1} \etF_k(i)\etF_k(i')} \rdbr*{\prod_{\set{u,v} \in E}\E\sqbr*{s_{\set{u, v}}\rdbr{i_u} s_{\set{u, v}}\rdbr{i'_u}s_{\set{u, v}}\rdbr{i_v} s_{\set{u, v}}\rdbr{i'_v}}} \\& \sum_{j\in J} \sum_{j'\in J} \E\sqbr*{\prod_{k=0}^{r-1} \1\rdbr*{H_k(i) = G_k(j)} \1\rdbr*{H_k(i') = G_k(j')}}
\end{align*}
The expected value of the sign functions is stated in the proof for Theorem~\ref{thm:dobra}. If we consider only the first disjunction, then we get:
\begin{align*}
    \E\sqbr{X}^2 \sum_{j\in J} \sum_{j'\in J} \E\sqbr*{\prod_{k=0}^{r-1} \1\rdbr*{H_k(i) = G_k(j)} \1\rdbr*{H_k(i') = G_k(j')}}
\end{align*}
which is equal to $\E\sqbr{X}^2$, because when $i_u=i_v\land i'_u = i'_v$, all graph components must have one or two distinct values each. In the case of one distinct value, $j_q = j'_q$ for that graph component $q$. Thus, the expected value per graph component is either $1/m^2$ or $\1(j_q = j'_q)/m$, making the sums over $J$ equal to 1. Now, let us consider everything but $\E\sqbr{X}^2$ in $\E\sqbr{X^2}$. There must be at least one occurrence of either $i_u = i'_u \neq i_v = i'_v$ or $i_u = i'_v \neq i_v = i'_u$. Either way, $j_q = j'_q$ for that graph component $q$ while there are still at least two distinct values. The sums over $J$ is thus $\leq 1/m$.
We then get that:
\begin{align*}
    \E\sqbr{X^2} ={}& \E\sqbr{X}^2 + Y \implies \Var(X) \leq \frac{1}{m}\E\sqbr{X}^2 + Y\\
    \Var(X) \leq{}& \frac{1}{m} \sum_{i \in I}\sum_{i' \in I} \etF_{0}(i) \cdots \etF_{r-1}(i)   \etF_{0}(i') \cdots \etF_{r-1}(i') \\&\prod_{\set{u,v} \in E}\1\rdbr*{i_u = i_v \land i'_u = i'_v} + \1\rdbr*{i_u = i'_u \neq i_v = i'_v} + \1\rdbr*{i_u = i'_v \neq i_v = i'_u}
\end{align*}
which is the same expression of the variance bound as the one for Theorem~\ref{thm:dobra}. The final steps of the variance bound are thus equivalent, resulting in the desired upper bound.
\end{proof}

\bibliographystyle{ACM-Reference-Format}
\bibliography{references}

\end{document}